\documentclass[]{aa}
\usepackage{natbib}
\usepackage{graphicx}

\bibpunct{(}{)}{;}{a}{}{,}

\newcommand\bb[1] {   \mbox{\boldmath{$#1$}}  }
\newcommand\del{\bb{\nabla}}
\newcommand\bcdot{\bb{\cdot}}
\newcommand\btimes{\bb{\times}}

\begin{document}

\title{Dust settling in local simulations of turbulent protoplanetary disks}
\author{ S\'ebastien Fromang \inst{1,2} and John Papaloizou \inst{2}}

\offprints{S.Fromang}

\institute{Astronomy Unit, Queen Mary, University of London, 
Mile End Road, London E1 4NS \and  Department of Applied Mathematics
and Theoretical Physics, University of Cambridge, Centre for
Mathematical Sciences, Wilberforce Road, Cambridge, CB3 0WA, UK \\ \email{S.Fromang@damtp.cam.ac.uk}}

\date{Accepted; Received; in original form;}

\label{firstpage}

\abstract{
In this paper, we study the effect of MHD turbulence on the
  dynamics of dust particles in protoplanetary disks. We vary the
  size of the particles and relate the dust evolution to the turbulent
  velocity fluctuations.
We performed numerical simulations  using two  Eulerian MHD codes,
  both based on finite difference techniques: ZEUS--3D and NIRVANA.
  These were local shearing box simulations incorporating vertical
  stratification. Both ideal and non ideal MHD simulations 
  with midplane dead zones were carried out.
  The codes  were extended to incorporate different models for the dust
  as an additional fluid component. Good agreement between results
  obtained using the different approaches was obtained.
The simulations show that a thin layer of very  small dust particles is
  diffusively spread over the full vertical extent of the disk. We show
  that a simple description obtained  using the diffusion equation
   with a  diffusion coefficient simply expressed  in terms of the velocity
  correlations accurately matches the results. 
   Dust settling starts to
  become apparent for particle sizes of the order of $1$ to $10$
  centimeters for which the gas begins to decouple 
  in a standard solar nebula model at $5.2AU.$
   However, for  particles which are $10$ centimeters in  size, complete
  settling toward a very thin  midplane layer is  prevented by 
  turbulent motions within the disk, even in the presence of a midplane dead
  zone of significant size.
These results indicate that, when present,  MHD turbulence affects dust
  dynamics in protoplanetary disks. We  find that the evolution and settling of 
  the dust can be accurately modelled  using  an advection diffusion
  equation that incorporates vertical settling. The value of the  diffusion coefficient can
  be calculated from the turbulent velocity field when that is known
  for a time of several local orbits.
}

\authorrunning{S.Fromang \&  J.Papaloizou}
\titlerunning{Dust settling in turbulent protoplanetary disks}
\maketitle

\section{Introduction}

Dust particles are very likely to be the basic building blocks
 that need to be assembled to  make planets.

In the  ``core--accretion'' model for the formation of giant planets
\citep{mizuno80,pollack96}, the growth of their solid  cores  
proceeds through the accretion of  objects ranging from
micron sized dust particles to planetesimals of radius $\sim 10 \ km$ 
to eventually reach  about $15 M_{\oplus}.$  Dust is also one of the main observational
tracers of the structure of protoplanetary accretion disks
\citep{adamsetal87,dalessioetal01}. A detailed knowledge of its 
dynamics is therefore needed both in order to make theoretical models and to give
the  best interpretation of the observations.

The key ingredient that drives the evolution of dust particles is the
drag force they
feel from the gas \citep{weidenschilling77}. Gas
must be present if a giant planet is to form subsequently. 
Drag  is responsible for
example for the radial migration of solid bodies toward the inner parts
of a  disk for which the pressure decreases radially outwards. 
The timescale for this migration is so rapid for
centimeter and meter size bodies that it has become a problem for
standard planet formation theories \citep{weidenschilling77},
a resolution of which might involve the particles accumulating
close to pressure maxima possibly at the center of vortices.

 Another
important aspect of dust dynamics is the tendency to settle towards
the midplane of the disk, which increases with particle size.
 Because of this , centimeter and
metre  sized bodies  accumulate close to the disk midplane. As
argued by \citet{goldreich&ward73}, a gravitational instability in the
dust sub--disk in triggered when the dust density becomes high enough
(see also \citeauthor{safronov69} \citeyear{safronov69} and, as
remarked in a note in proof by \citeauthor{goldreich&ward73}
\citeyear{goldreich&ward73}, \citeauthor{gurevich&lebedin50}
\citeyear{gurevich&lebedin50}).
At this point,  larger bodies known as planetesimals form.

 However, because  the
pressure force acts only on the gas , a velocity shear between the dust sub--disk and
the upper layers of the gas disk develops. 
This   can lead to the so called
``shear--induced'' turbulence that may be able to mix the dust layer
enough to prevent enough gravitational settling  to satisfy the condition
for gravitational  instability
\citep{garaudetal04,gomez&ostriker05}. However, the onset of this
shear induced turbulence depends on the vertical profile of the dust
sub--disk which
is quite difficult to calculate. Indeed, this is determined  as the result of the
interplay between various physical processes, such as dust coagulation,
Brownian motions and turbulent mixing 
\citep{dullemond&dominik05}. The
purpose of this paper is to investigate the behaviour of the dust disk when just one of these
operates, namely  MHD turbulence, from first principles.

Turbulence in protoplanetary disks is believed to be the result of the
magnetorotational instability (MRI). More than a decade of analytical and
numerical work (see a review by \citeauthor{balbus&hawley98}
\citeyear{balbus&hawley98}) has shown that this is capable of leading
to sustained turbulence,  and magnetic field with associated stresses that transport angular
momentum outwards. More and more of its properties have been analysed
thanks to the increased computational power currently available.
 However, little  work has  focused on the effects of MHD turbulence 
on dust dynamics. \citet{carballidoetal05} and
\citet{johansen&klahr05} analysed the {\it radial} diffusion of small
particles and calculated the  effective Schmidt number, which is the
ratio between the anomalous turbulent viscosity and the anomalous diffusion
coefficient. They found different results which could be
due to their different initial setup. \citet{johansenetal05} and
\citet{fromang&nelson05} studied the {\it radial}  migration of centimeter
and meter size bodies in turbulent disk models and found that dust
trapping in local pressure
maxima could be an important confining  effect. Very recently, \citet{turneretal06}
investigated the dispersion of a passive scalar in stratified disk
models by releasing particles at high altitude in the disk. They
interpreted the effect of MHD turbulence in terms of a damped wave
equation. However their analysis was restricted to infinitely small
particles. They did not consider the vertical settling that begins to  occur 
for larger particles. Thus in their case stratification is only significant
for the gas, being felt by the dust particles only through their strong coupling to the gas.

A first attempt to  study vertical settling numerically  was made
through global SPH calculations by \citet{barriereetal05}. They confirmed
earlier analytical results regarding settling in a quiescent disk
\citep{garaud&lin04}. However, the
question of the effect of MHD turbulence was left unanswered. In this
paper, we tackle this problem by means of local MHD simulations that use
Eulerian finite difference codes. Our goal is to relate the dust
dynamics to  the  properties of the turbulence. and also to characterise the amount
of settling as a function of the particle size. Because of the complexity
of this problem, we neglect dust coagulation and only consider 
populations of particles with a single specified size.

The plan of the paper is as follows: in
section~\ref{dust_gas_interaction}, we review the characteristics of
dust--gas interaction and describe a simple model to account for the
dispersion of a passive
scalar in a turbulent medium with random velocity field. 
In section~\ref{num_methods}, we present
the numerical methods we used together with the basic disk model
which exhibits sustained MHD turbulence. We go on
 to  study of the diffusion of very small dust particles
in section~\ref{small_part}. Their evolution is analysed in the
context of the simple diffusion  model discussed in
section~\ref{dust_gas_interaction}. We then show in
section~\ref{large_part} that larger dust particles
start to decouple from the gas and settle toward the midplane of the
disk. In this case the evolution  is found to
be well described using an advection diffusion equation incorporating settling
together with the same diffusion coefficient that applied to the very small particles.
 In section~\ref{dead_zone}, we investigate the
effect a dead zone is likely to have on these processes by setting
up non--ideal MHD calculations that result in a significant
midplane centered dead zone. However, gas motions are effective
at preventing complete settling for $10 cm$ sized
particles in this case too.
 Finally, we discuss our results 
 in section~\ref{discussion}.
\section{Basic equations}
\label{dust_gas_interaction}

We  adopt a two fluid model of the dust and gas circulating in a
protoplanetary accretion disk.
The basic equations are those of MHD applied to the gas component together
with those of hydrodynamics applied to the dust component. The model allows
 the  two fluids to
have different flow velocities and consequently
exchange momentum through drag forces. 

The basic equations for the gas  component
are the continuity, momentum conservation  and induction 
equations. In a frame rotating with angular velocity ${\bb{\Omega}} = \Omega {\hat {\bf k}},$ with
 ${\hat {\bf k}}$ being the unit vector in the fixed direction of rotation, here called the vertical direction and $\Omega$ being the magnitude of the angular velocity, these take the form 
\begin{eqnarray}
\frac{\partial \rho}{\partial t} + \del \bcdot (\rho \bb{v})  &=&  0 \,\label{contg} , \\
 \frac{\partial \bb{v}}{\partial t} + ( \bb{v} \bcdot \del )
\bb{v}   + 2
\bb{\Omega} \times \bb{v}   &=&  - \frac{1}{\rho} \del P + \frac{1}{4\pi\rho} (\del \btimes \bb{B})
\btimes \bb{B} \nonumber \\
&& \hspace{1.05cm}  
   -\nabla \Phi - \frac{\rho_d}{\rho }{\bb{F_{drag}}\over{m_p}}  \,\label{mog} , \\
\frac{\partial \bb{B}}{\partial t}  &=&  \del \btimes ( \bb{v} \btimes
\bb{B}  - \eta \del \btimes \bb{B} ) \, \label{induct} .
\label{shearing_sheet_eq}
\end{eqnarray}
Here, $P$ is the gas pressure, $\rho$ is the gas density, $\Phi$ is sum of the gravitational
potential, here due to a central mass and the centrifugal potential
 $ -\Omega^2 |{\bf r}\btimes {\hat {\bf k}}|^2/2 $ with ${\bf r}$ 
being the position vector, ${\bb{v}}$ is the gas velocity
 and $\bb{B}$
denotes the magnetic field. The resistivity of the gas  $,\eta,$  
is taken to be nonzero only in
section~\ref{dead_zone} where we investigate the effect of non--ideal
MHD. The last term in equation (\ref{mog}) is the acceleration
of the gas due to  drag on the dust component. The drag force acting on 
a single dust particle of mass $m_p$ is $\bb{F_{drag}}$ (see below) 
and $\rho_d$ is the dust density.

\noindent The equations for the dust  are

\begin{eqnarray}
\frac{\partial \rho_d}{\partial t} + \del \bcdot (\rho_d \bb{v}_d) 
 &=&  0 \,\label{contd} , \\
 \frac{\partial \bb{v}_d}{\partial t} + ( \bb{v}_d \bcdot \del )
\bb{v}_d   + 2
\bb{\Omega} \times \bb{v}_d   &=&  -\nabla \Phi + {\bb{F_{drag}}\over{m_p}}  
  \,\label{mod} , 
\label{shearing_sheet_eqdust}
\end{eqnarray}
 with $\bb{v}_d$ being the velocity of the dust component. 

\subsection{The drag force}

 The  dust interacts
with the gas  through drag. 
In this paper, we consider only dust particles that are small enough in
size that they   are in the
Epstein regime \citep{weidenschilling77}. The drag force
$\bb{F_{drag}}$ acting on a single particle then takes the form
\begin{equation}
\bb{F_{drag}}=\frac{m_p}{\tau_s}(\bb{v}-\bb{v}_d) \, .
\end{equation}
  This drag force is proportional to $\bb{v} - \bb{v}_d$ which is 
 the  relative velocity  between 
the gas  and dust  components, and $\tau_s$, the dust 
stopping time, defined by
\begin{equation}
\tau_s=\frac{\rho_{s} a}{\rho c_s} \, ,
\end{equation}
where  $a$   is the dust particle radius, $\rho_s$ is the dust particle density
and $c_s$ is the  local speed
of sound which, as we shall take an isothermal equation of state, is  independent
of height.

When $\Omega$ is taken to be the Keplerian angular velocity
at some disk radius,
the size of the particles there
can be measured in terms of the dimensionless parameter
$\Omega\tau_s$ through the relation 
\begin{equation}
a=\left( \frac{\rho}{\rho_s} \right) ( \Omega \tau_s ) H \,\label{size} ,
\end{equation}
where $H = c_s/\Omega $ is the  disk scale height. Following \citet{johansenetal04},
 when we  take $\rho_s/\rho=10^{10}$ and $H=10^{12}$ cm, equation (\ref{size})
gives 
\begin{equation}
a=100 \left( \Omega \tau_s \right) \, \textrm{cm} \, .
\label{size_part}
\end{equation}
The above relation is also  appropriate for the standard minimum
mass solar nebula at $5.2 AU.$

Using an approach based on Boltzmann averaging,
\citet{garaudetal04}  argued that the  evolution of the dust particle
distribution  can be 
accurately determined by modelling it as a  pressureless fluid  as long as the
dimensionless parameter $\Omega \tau_s$ is   less than unity. This is
the case for the  dust particles   over most of the
disk domains considered in this
paper.  Accordingly we adopt the two fluid description  in preference
to, for example a much more computationally demanding  N--body
approach. We also assume throughout this paper that the dust particles
remain  electrically neutral so that they do not   react to the
magnetic field. 

\subsection{ Single fluid with advection diffusion equation and settling}
\label{single_fluid}
We here note that in the limit $\tau_s \rightarrow 0,$ the
two fluid description can be reduced to one of a single fluid
with MHD together with an advection diffusion equation describing the evolution of the 
dust. To see this we  introduce a density $\overline{\rho}$ and a velocity
 $\overline{\bb{v}}$ associated with the combined fluid.
These are defined through

 \begin{equation}
\overline{\rho} = \rho + \rho_d
\label{combrho}
\end{equation}
and
 \begin{equation}
\overline{\rho} \, {\overline {\bb{v}}} = \rho{\bb{v}}+ \rho_d{\bb{v}_d}.
\label{combvel}
\end{equation}
 Adding the continuity equations (\ref{contg}) and (\ref{contd})
and  eliminating the drag force  by combining  equations (\ref{mog}) and (\ref{mod})
and  subsequently 
letting $\tau_s \rightarrow 0$ while assuming $\bb{v}-\bb{v}_d = O(\tau_s)$
gives

\begin{eqnarray}
{{\partial {\overline \rho}}\over {\partial t}} + \del \bcdot
({\overline \rho} \, 
{\overline {\bb{v}}})  &=&  0 \,  \label{contt} , \\
 {\overline \rho}\left(\frac{\partial {\overline {\bb{v}}}}{\partial t} +
 ( {\overline {\bb{v}}} \bcdot \del )
{\overline {\bb{v}}}   + 2
{\bb{\Omega}}\times {\overline {\bb{v}}}\right)   &=&  -  \del P 
+ \frac{1}{4\pi} (\del \btimes \bb{B})
\btimes \bb{B} \nonumber \\
&& \hspace{1.05cm} 
   -{\overline \rho}   \nabla \Phi  \,\label{mot} . 
\label{shearing_sheet_aveq}
\end{eqnarray}

Correct to first order in $\tau_s,$ we may set ${\bb{v}}_d = {\overline {\bb{v}}}$
in the left hand side of equation (\ref{mod}) and then use that equation  to find ${\bb{v}}_d$
located on the right hand side.  Substituting this into 
equation (\ref{contd}) and after   making use of equation (\ref{mot})
gives rise to the  equation 

\begin{equation}
\frac{\partial \rho_d}{\partial t} + \del \bcdot (\rho_d {\bb{u}})
 =  0 \,\label{contds} 
\end{equation}
with
\begin{equation}
 {\bb{u}}
 =   {\overline{\bb{v}}}  + {\overline{\bb{v}}_s},
\end{equation}
where 
\begin{equation}
 {\overline{\bb{v}}_s}=
 -{\rho\tau_s\over {\overline \rho}^2}\left(-  \del P
+ \frac{1}{4\pi} (\del \btimes \bb{B})
\btimes \bb{B}\right) \,\label{dvel}
\label{advectdiff}
\end{equation}

Using the above in equation (\ref{contds})  and setting $\rho_d = X{\overline \rho},$
$\rho =(1- X){\overline \rho},$ results in the following equation for the dust mass
fraction $X$

\begin{eqnarray}
&& {\overline \rho}\left( {{\partial X}\over {\partial t}} + 
{\overline {\bb{v}}}\cdot\nabla X\right) + \nonumber \\
&& \nabla\cdot\left[ X\rho\tau_s\left({(1-X)\nabla({\overline \rho}c_s^2)\over {\overline \rho}} +  
\frac {(\del \btimes \bb{B})\btimes \bb{B}}{4\pi{\overline \rho}}
\right) \right] \nonumber \\
&& =  
\nabla\cdot (\rho c_s^2\tau_s X\nabla X),\label{advde}
\end{eqnarray}
This can be interpreted as an advection diffusion equation
where the advection velocity ${\bb{u}}_a$ is the sum of the mean fluid velocity
and a settling velocity such that
\begin{equation}
 {\bb{u}}_a
 =   {\overline{\bb{v}}} +{\rho\tau_s \over {\overline \rho}}
\left({(1-X)\nabla({\overline \rho}c_s^2)\over {\overline \rho}} +
\frac {(\del \btimes \bb{B})\btimes \bb{B}}{4\pi{\overline \rho}}
\right).
\end{equation}
We remark that for small $X$ and no magnetic field,
the vertical component of the settling velocity is $-g\tau_s$
with $g$ being the vertical acceleration due to gravity.
In addition there is a small diffusion coefficient arising through 
the second derivatives of $X$ with respect to the
coordinates such that
${\cal D} = \rho X  c_s^2\tau_s.$ In practice, when turbulence is present,
this is overwhelmed
by the effect of the mean fluid velocity, that  being particularly the case
when $X$ is small.

\subsection{A simple theory for dust diffusion}
\label{simple_th}

The above analysis suggests that the dust mass fraction is advected
as a passive scalar. When the gas is turbulent  part of the velocity
field will be turbulent. The characteristic correlation
time associated with the turbulence is expected to be a fraction 
of an orbital period (see below), whereas the underlying settling
takes place on a significantly longer timescale $\sim 1/(\Omega^2 \tau_s).$
Thus we separate the effects of turbulent diffusion and settling
    and 
consider the simplest case of   the dispersion of a
passive scalar in steady state, homogeneous and isotropic
turbulence. This means that effects related to the imperfect coupling
between dust and gas such as turbulent enhancement of grain--grain
collisions are neglected \citep{volketal80,morfill85}. It is well
known \citep{taylor21,batchelor49} that the mean
square  displacement of the scalar from its location
at some initial time  can be expressed in terms of the fluid
velocity fluctuations. We adopt a Lagrangian approach and for simplicity
consider  only the vertical  $z$--direction. Extension to consider the
other coordinate directions is straightforward but uninformative in our case. 

\noindent Let
$z(z_0,t)$ be the position of a particle which was located at $z=z_0$
 at $t=0$. Its displacement $Z(z_0,t)$ is  given by 
\begin{equation}
Z(z_0,t) = z(z_0,t)-z_0 = \int_0^t U(z_0,t') dt' \, ,
\end{equation}
where $U(z_0,t)$ is the vertical component of
the  velocity of the particle as  defined in a Lagrangian
sense as seen  along its
path. It relates to its Eulerian equivalent $v_z(z,t)$ through
  $U(z_0,t)=v_z(z,t)$. We are interested in the mean square deviation of 
particles from their initial positions. Using the fact that $U(z_0,t)$
and $Z(z_0,t)$ are related by a time integral, the derivative of
$Z^2(z_0,t)$ with respect to $t$ can be expressed in terms of the
velocities as: 
\begin{eqnarray}
\frac{\partial Z^2(z_0,t)}{\partial t}&=&2\int_0^t U(z_0,t')U(z_0,t)dt' \nonumber \\
&=&2\int_0^t v_z(z(z_0,t'),t')v_z(z(z_0,t),t)dt' \, .
\end{eqnarray}

 To obtain a mean square particle displacement
we take  an ensemble average  over many particle realisations,
 indicated by angular brackets, to 
 obtain
\begin{eqnarray}
& & \frac{d <Z^2(z_0,t)>}{dt} = 2\int_0^t S_{zz}(t,\tau) d\tau = \nonumber \\
&& 2\int_0^t <v_z(z(z_0,t-\tau),t-\tau)v_z(z(z_0,t),t)> d\tau \, ,
\label{Szz_lagrangint}
\end{eqnarray}
where $S_{zz}(t,\tau)$ is the velocity correlation function  which
depends only on the  properties of the turbulence. For
turbulence in a statistically steady state, as there is no preferred time,
 it should depend only
on  the time difference $t-t' = \tau.$ Thus any value of $t$ may be 
adopted in $S_{zz}(t,\tau)$  so that
\begin{equation}
S_{zz}(t,\tau)=S_{zz}(t=0,\tau)= S_{zz}(t=\tau,\tau) \, ,
\label{Szz_lagrang}
\end{equation}
which finally writes
\begin{equation}
S_{zz}(t,\tau)=<v_z(z(z_0,\tau),\tau)v_z(z_0,0)> \, .
\end{equation}
When $\tau  \rightarrow 0$, $S_{zz}$   becomes $<v_z^2>$, i.e. it is a
measure of the mean square turbulent velocity fluctuations.
 In the opposite limit , when $\tau \rightarrow \infty,$ 
 $S_{zz}$ tends to zero as the velocities become uncorrelated.

 The  calculation of the velocity correlation described above
is greatly simplified if the  integral in (\ref{Szz_lagrangint})
 can be performed at a fixed location using Eulerian velocity components,
such that $v_z(z(z_0,\tau),\tau)$  is replaced by $v_z(z_0,\tau)$ with
$z_0$ being fixed.
This is possible  when the distance moved by a particle during the
turbulence correlation time is small compared to the characteristic
scale of the turbulence \citep{biferaleetal95} as would always be
the case if the turbulent velocities were reduced in magnitude
while retaining their spatial and temporal characteristics. This situation
seems to be a reasonable approximation in the case of the MRI where
the turbulence is subsonic  and we note that it would be even better
in dead zones where the turbulent velocity fluctuations are reduced in
magnitude. In this case  $S_{zz}$ simply scales down as the square of
the velocity amplitude. Furthermore for   the turbulence we consider,
there is no preferred location or particles
 in the above ensemble averages  and all fixed locations  can be assumed
to get complete information about all particle realisations.   
 Then it is reasonable to replace the
ensemble average over all possible particle realisations as a function 
of time,  by one over all  possible  fixed locations using the
velocities there at that time. Therefore, from now on, we  adopt
\begin{equation}
S_{zz}(\tau)=<v_z(z,\tau)v_z(z,0)> \, ,
\label{Szz_func}
\end{equation}
where,  the ensemble average  is over 
fixed locations $z$ and over different realisations
obtained from  a set of models. We also define the quantity
$D_T(\tau)$ to be
\begin{equation}
D_T(\tau)=\int_{0}^{\tau} S_{zz}(\tau') d\tau' \, .
\label{diff_coeff_func}
\end{equation}
\subsection{Diffusion coefficient}
 As we shall see, for large
$\tau,$ the quantity $D(\tau)$ defines an effective diffusion
coefficient. For small $\tau$, 
$S_{zz}$   takes on a finite value.   Thus we expect $D_T(\tau) \propto 
\tau$. This is  known as the ballistic regime \citep{taylor21}
\begin{equation}
D_T(\tau) \sim <v_z^2> \tau \hspace{0.5cm} \textrm{(for small } \tau
\textrm{).}
\label{small_tau}
\end{equation}
When $\tau$ becomes large, $S_{zz}$ goes to zero, so we can postulate  that
$D(\tau)$ goes to a finite limit,  meaning that $<Z^2>$ would scale
like $\tau$. This is the diffusive regime. If $\tau_{corr}$ is a
typical correlation timescale for the turbulence, then we expect, from
simple dimensional analysis
\begin{equation}
D_T(\tau) \sim <v_z^2> \tau_{corr} \hspace{0.5cm} \textrm{(for large }
\tau \textrm{).}
\label{large_tau}
\end{equation}
However, at this point we insert a cautionary note.
Strict convergence as $\tau \rightarrow \infty$ would require
ensemble averaging over an infinite number of realisations.
In practical  cases these will be finite in number
and evaluating $D_T(\tau)$ for increasing $\tau$  corresponds to working out the average of 
a finite number of random walk displacements for  increasing
numbers of steps.  Although this might appear reasonable for modest $\tau,$  it  would 
not be expected to show strict convergence for very large $\tau$
in practical cases. For this reason a time span of $8$ orbits equivalent
to about $50$ correlation times is adopted below.

In  this context we comment that the modelling of dust spreading
using a diffusion coefficient obtained by the above procedure,
on account of the averaging involved, of necessity only describes evolution
occurring on time scales significantly  longer than the correlation time associated
with the turbulence. Behaviour occurring on the correlation time scale or faster cannot
be meaningfully considered in this approach.

\section{Numerical Methods}
\label{num_methods}

\subsection{The algorithms}

In this paper, we use two well known finite difference codes: ZEUS-3D
\citep{stone&norman92a,stone&norman92b} and NIRVANA
\citep{ziegler&yorke97}. Both are  
used here to  solve the
 MHD equations in the shearing sheet approximation
\citep{goldreich&lyndenbell65}, including vertical stratification.
In this approximation a small region of the disk is considered
in the neighbourhood of some point rotating in circular
orbit  with the local Keplerian angular 
velocity. Local Cartesian coordinates are used with the $x$ axis 
 along the line connecting the origin to the central mass but pointing away from it, and
the $y$ axis directed in the direction of orbital motion.
When dust is absent the equations solved are  (\ref{contg} -- \ref{induct})
with  $ {\bb{F_{drag}}}=0$ adapted to the shearing sheet
by taking
\begin{equation}
\nabla \Phi = -3\Omega^2 x{\hat {\bf i}} +\Omega^2z{\hat {\bf k}},
\end{equation}
with ${\hat {\bf i}}$ being the unit vector in the $x$ direction.
The evolution of the magnetic field is calculated using the MOC-CT
algorithm \citep{hawley&stone95}, such that the solenoidality
constraint on the magnetic field is satisfied at all times.

In the two codes, we employ two different algorithms to describe the
evolution of the dust particles that  are the  main  focus of this
paper. ZEUS--3D was extended to describe the dust particle evolution  by means of a
second, pressureless fluid \citep{fromang&nelson05} using equations
(\ref{contd}-\ref{mod}) together with appropriate drag forces in the low $X$ limit.
Because of the
short stopping time of the dust particles, the effect
of these  is computed implicitly. 

On the other hand, NIRVANA solved the MHD equations (\ref{contt}-\ref{mot})
together with (\ref{induct}). Consistent with the low $X$ limit the mean
flow velocity ${\overline {{\bb{v}}}}$  was used in (\ref{induct}).
The evolution of the dust mass fraction was calculated by solving
equation (\ref{advde}) but with the use of an advection
velocity limiter. This was applied such as to restrict the settling velocity
$ {{\bb{u}}}_a - {{\bb{v}}}$ to be less than $0.1c_s$ in magnitude.
This is required in the low density upper layers where dust gas coupling becomes weak and where the theoretical description breaks down.
However, because it applies only in the upper layers, calculations
of settling dust are insensitive to this cut off that is required for numerical reasons.

\subsection{The model properties}
\label{model_prop}

In this section, we describe the underlying disk model we used and
highlight some of the properties of the MHD turbulence that are of
importance for the dynamics of the dust that we go on to study.

The initial disk setup in the absence of dust is very similar to the model of
\citet{stoneetal96}. The parameters are those of the
standard shearing box. The equation of state for the gas is isothermal, $P=\rho
c_s^2$. Because of the vertical stratification, the initial density
distribution is given by:
\begin{equation}
\rho \equiv {\overline \rho} =\rho_0 e^{-z^2/2H^2} \, ,
\end{equation}
where $\rho_0$ is the midplane density  and $H=c_s/\Omega$ is the disk
 scale height. The Cartesian box extends over the domains
 $[-H/2,H/2]$ in $x$, $[0,2\pi H]$ in $y$ and $[-3H,3H]$ in $z.$ We use
the standard  periodic boundary conditions in shearing coordinates
in  $x$
\citep{hawleyetal95,balbus&hawley98} and periodic boundary conditions
in $y$. One needs to be careful with the vertical  boundary conditions.
Because of
the vertical density stratification, in particular
the non vanishing vertical component 
of the gravitational force on the boundary,  using simple periodic boundary
conditions in that direction leads to  unphysical  fluctuations generated in
the box. 

In ZEUS--3D, we therefore  made the vertical gravitational force 
continuous at the boundary by
writing the gravitational potential as
\begin{equation}
V_{grav}=\textrm{max}\left(-\frac{1}{2} \Omega^2 (z^2-H_0^2), 0\right) \, .
\end{equation}
In practice, this  means taking   vertical gravity into account only in the
region $z \in [-H_0,H_0]$. We found that taking $H_0=2.4 H$ produces
satisfactory results.

In NIRVANA, we  further checked for the existence
of numerically generated fluctuations
by applying  the additional procedure of making the vertical
gravitational  acceleration continuous throughout.
To do this the usual  acceleration was applied for  $z \in [-2.25H,2.25H].$
This was then  reduced to zero in $\pm[2.25H ,2.4 H]$  
by means of linear interpolation and
set to zero for $|z|> 2.4H.$ As indicated below, similar results were
obtained with both codes.
  
In both cases the computational box is initially threaded by a magnetic field
with  zero net flux, being of the form 
\begin{equation}
B_z=B_0 \sin (2\pi x/H) \, .
\end{equation}
$B_0$ is calculated such that $\beta$, the ratio between the thermal
pressure and the magnetic pressure is initially $100.$ 
 Finally, the numerical resolution is
$(N_x,N_y,N_z)=(32,100,192)$ for ZEUS and $( 36, 100, 196)$ for NIRVANA. 
 Models with no dust are initiated by imposing a
 small random  velocity component  and subsequently
 run for $\sim 100$ orbits.

\begin{figure}
\begin{center}
\includegraphics[scale=0.5]{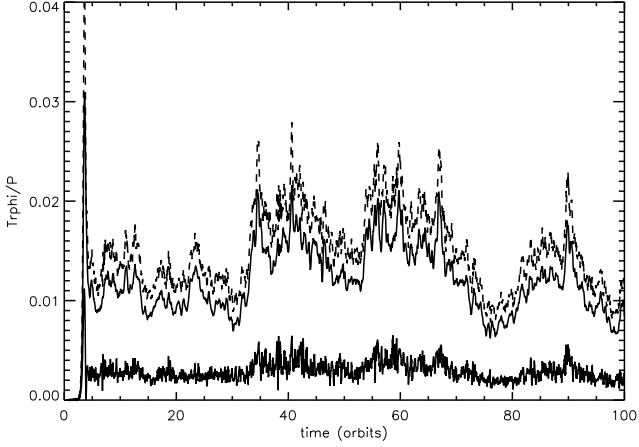}
\caption{Time history of the total stress ({\it dashed line}), the sum of
the Maxwell ({\it upper solid line}) and the Reynolds stresses ({\it
  lower solid line}) obtained with ZEUS--3D. All stresses are
normalised by the midplane
pressure. After the linear growth of the MRI, the total stress reaches
a nonzero quasi--steady state, showing that the turbulence is sustained.}
\label{stresses_history}
\end{center}
\end{figure}

\begin{figure}
\begin{center}
\includegraphics[scale=0.5]{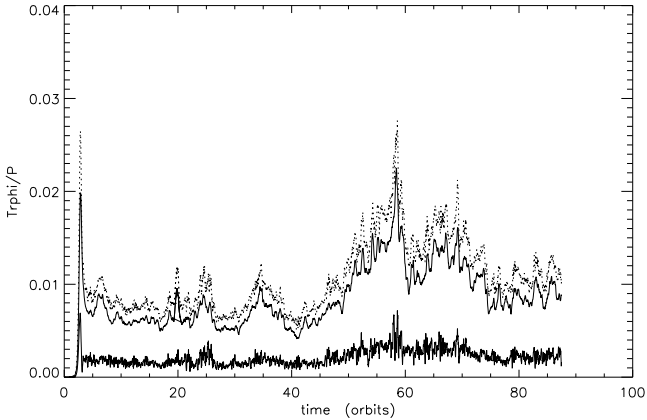}
\caption{As in Figure \ref{stresses_history} but obtained with NIRVANA.}
\label{stresses_historyN}
\end{center}
\end{figure}

\begin{figure}
\begin{center}
\includegraphics[scale=0.06]{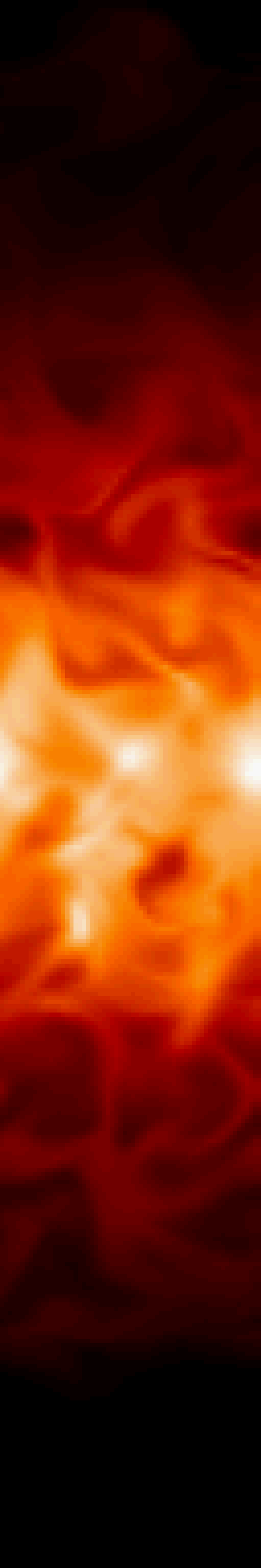}
\caption{Gas distribution in the (r,z) plane. Density fluctuations
  across the disk are clearly visible, showing that the entire disk
  is turbulent at this stage.}
\label{gas_distrib}
\end{center}
\end{figure}

\begin{figure}
\begin{center}
\includegraphics[scale=0.5]{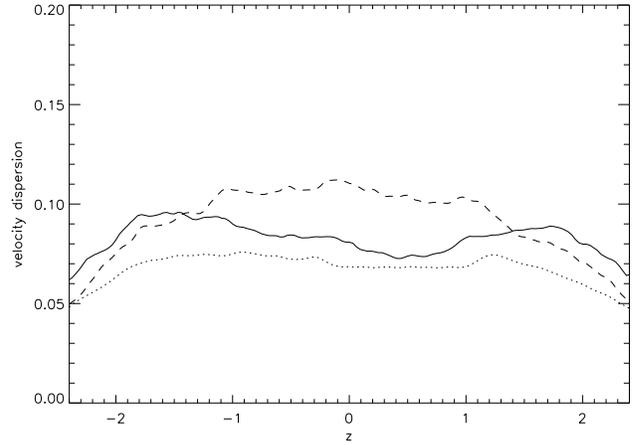}
\caption{Velocity dispersion as a function of height, normalised by
  the sound speed. The three curves corresponds to the radial ({\it
  dashed line}), azimuthal ({\it solid line}) and vertical ({\it dotted
  line}) velocity dispersions.}
\label{vel_disp}
\end{center}
\end{figure}

We describe here  simulation results, focusing on those
obtained with ZEUS--3D. The results obtained using NIRVANA were
very similar and accordingly only a few need to be illustrated here.
 As expected, the initial set up is unstable to the MRI. 
Its early growth is illustrated in
figure~\ref{stresses_history}. Corresponding results obtained with NIRVANA
are plotted in Figure~\ref{stresses_historyN}. Both figures show the
time history of the
volume averaged Maxwell stress ({\it upper solid line}) and Reynolds
stress ({\it lower solid line}), normalised by the average  midplane
pressure $P_0$. They are respectively defined as
\begin{eqnarray}
T_{r\phi}^{Max}&=&\int\!\!\!\int\!\!\!\int -{B_x B_y\over 4\pi} d\tau \, , \\
T_{r\phi}^{Rey}&=&\int\!\!\!\int\!\!\!\int \rho (v_x-\bar{v_x})(v_y-\bar{v_y}) d\tau
\, ,
\end{eqnarray}
where $\bar{v_x}$ and $\bar{v_y}$ are respectively the vertically and
azimuthally averaged radial and azimuthal velocities and $d\tau$ denotes
the element of volume. The dashed line in
figure~\ref{stresses_history} is the sum of the Maxwell and Reynolds
stresses: 
\begin{equation}
\alpha=\frac{T_{r\phi}^{Max}+T_{r\phi}^{Rey}}{P_0}
\end{equation}
After a few orbits, it peaks at a maximum as the flow breaks
down into turbulence and attains a saturated state. For the remaining part
of the simulation, $\alpha$  varies noisily  between
$0.006$ and $0.03$,
consistent with previous studies \citep{stoneetal96}. We remark that
there are quite large fluctuations in the stresses. At the
present time, their origin is not clear.  The
origin of these large variations in our stratified zero net flux models
 should be addressed in the future
using higher resolution calculations. In any
case, the results show clear evidence that MHD turbulence is sustained
for a very long time.  The state of the disk is illustrated in
figure~\ref{gas_distrib}, where we plot a typical snapshot of the
density in the (r,z) plane. Turbulent density fluctuations superposed
on the overall vertical stratification are obvious from this figure.

In section~\ref{simple_th}, we argued that the diffusion of small dust
particles depends on velocity fluctuations in the disk. In
figure~\ref{vel_disp}, we plot their vertical profile normalised by the
sound speed. The dashed
curve corresponds to the radial velocity fluctuations, while the
azimuthal and vertical fluctuations are respectively given by the solid
and dotted curves. The midplane values given by this plot are
\begin{eqnarray}
(\delta v_x^2)^{1/2}=0.11 c_s \nonumber \, , \, \, \, 
(\delta v_y^2)^{1/2}=0.08 c_s \nonumber \, , \, \, \, 
(\delta v_z^2)^{1/2}=0.07 c_s \nonumber \, .
\end{eqnarray}
These values are consistent with numbers previously reported in the
literature \citep{stoneetal96}.

\section{Vertical diffusion of small particles}
\label{small_part}

In this section, we present the results we obtained  for very small
particles using ZEUS--3D. For a given disk model, simulations of dust
evolution are defined by the
value of  $\Omega \tau_s$. This is actually  a function of gas density.
To  obtain a fixed parameter defining a given simulation, $\Omega\tau_s$ is
evaluated using the initial uniform midplane disk gas density. Where
this is not stated explicitly it should be taken as read. We focus here
on particles for which $\Omega \tau_s=10^{-5}$ in the midplane of the
initial disk. Equation~(\ref{size_part}) indicates that this
corresponds to micron size particles. This small value of $\Omega
\tau_s$ means that the dust particles are very well coupled
to the gas and behave almost like a passive scalar.

\begin{figure}
\begin{center}
\includegraphics[scale=0.15]{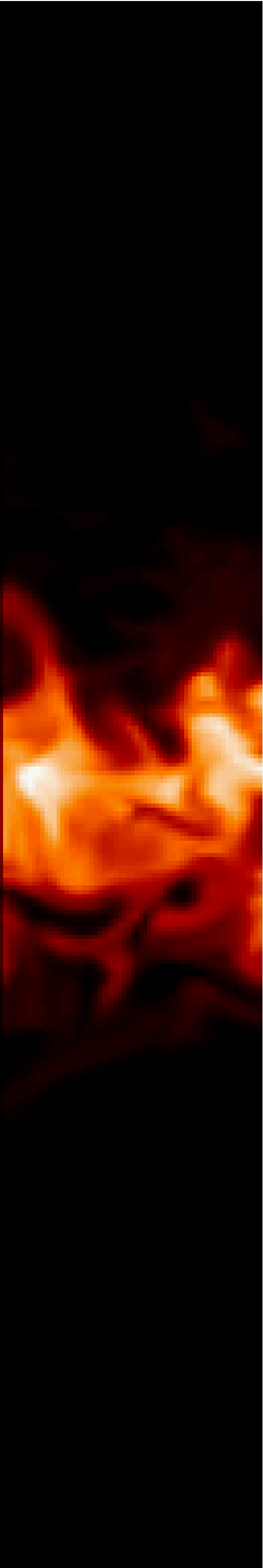}
\includegraphics[scale=0.15]{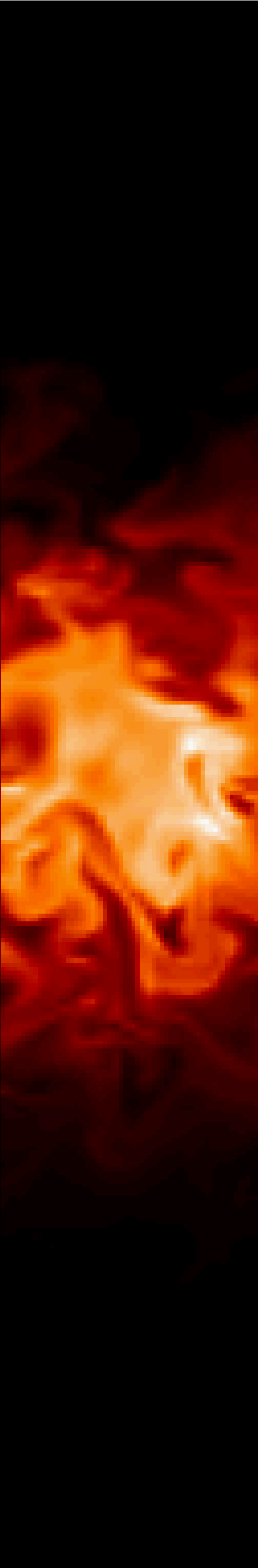}
\includegraphics[scale=0.15]{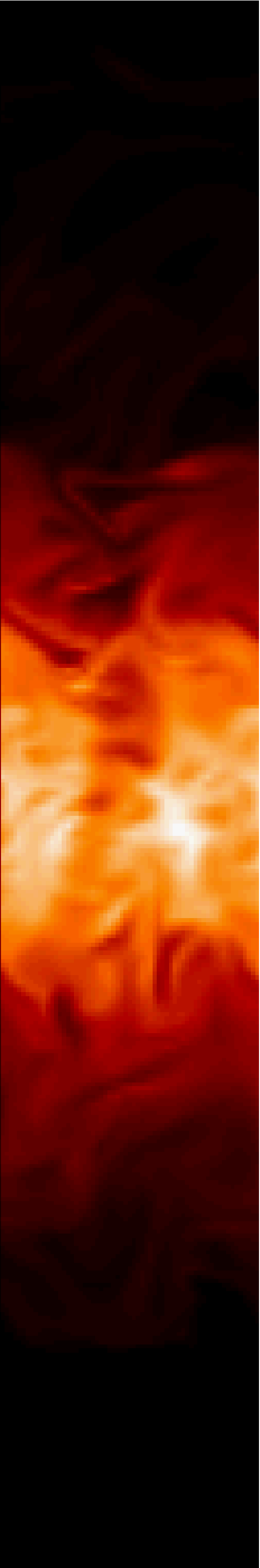}
\includegraphics[scale=0.15]{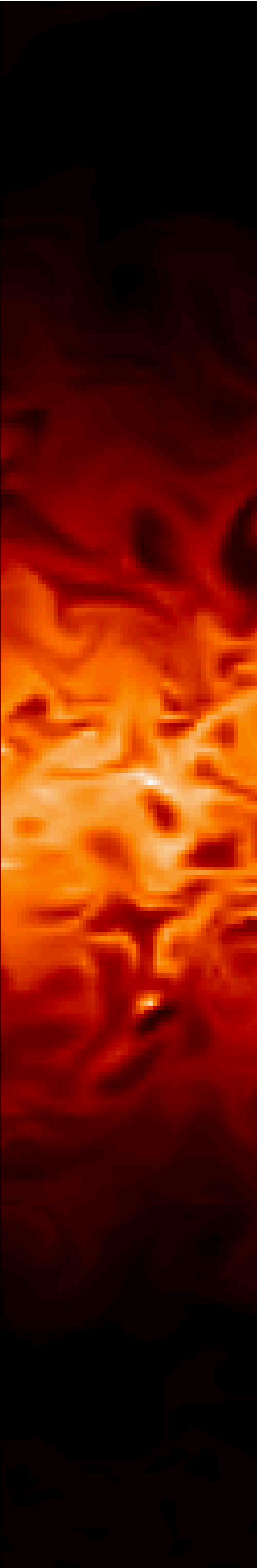}
\hspace{0.5cm}
\includegraphics[scale=0.15]{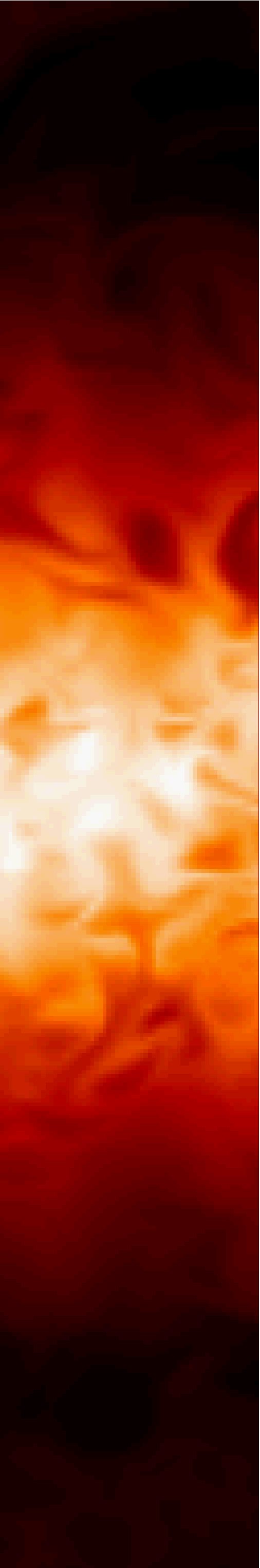}
\caption{The first four snapshots show the dust distribution in the
  (r,z) plane obtained with ZEUS--3D at times $t=3$, $5$, $10$ and
$15$ orbits after the dust is introduced ({\it from left to
  right}). The last snapshot on the right represents the gas
distribution at $t=15$ orbits. At that time, the micron--sized dust
particles are well mixed in the entire disk. The dust density
distribution is also seen to be well correlated with the gas density.}
\label{micron_dust}
\end{center}
\end{figure}

The dust is  initially distributed in a thin layer around the disk
midplane. Initially, the vertical profile of $\rho_d$ is taken to be a
Gaussian with a thickness $H_{d}=0.2H$. Under the effects of MHD turbulence,
this small layer broadens with time. We found that its precise
evolution depends largely on the particular time at
which the dust particles are released in the disk. In order to improve
the statistics, we computed nine simulations,
by restarting the disk model described in section~\ref{model_prop} at
times $t=40$, $45$, $50$, $55$, $60$, $65$, $70$, $75$ and $80$ orbits.

Figure~\ref{micron_dust} illustrates the typical evolution of such a
model. From left to right, the first four snapshots on the left show
the dust density distribution in the (r,z) plane at time $t=3$, $5$,
$10$ and $15$ orbits (measured after the dust has been
introduced). The last snapshot shows the corresponding gas distribution at
$t=15$ orbits. The dust is seen to spread quite rapidly. At $15$
orbits, it is almost filling the entire vertical extend of
the disk. At later times, this distribution does not change
qualitatively. By comparing the dust and the gas distribution at
$t=15$ orbits, one can also see that they are very well correlated. We
also found that the initial distribution of the dust has very little
effect on this final state: starting with an initial dust distribution
such that the gas--to--dust ratio is uniform gives an end product
almost indistinguishable from the last two
snapshots of figure~\ref{micron_dust}.

\begin{figure}
\begin{center}
\includegraphics[scale=0.6]{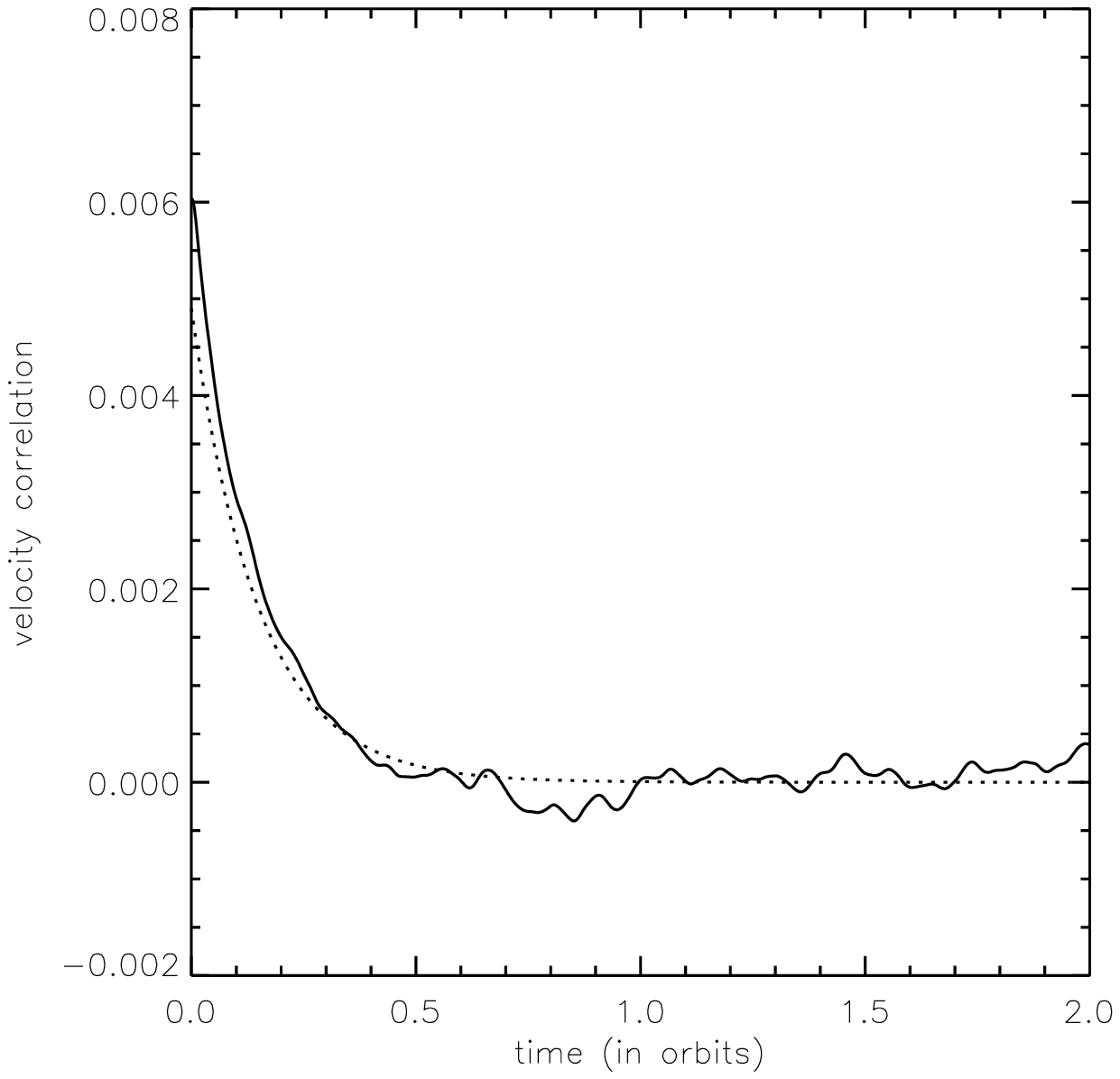}
\caption{Time history of the function $S_{zz}(\tau)$, averaged between the nine
  models that were run with ZEUS--3D. The solid line show the function
  as it is extracted from the
  models while the dotted line represents the function
  $S_0 \exp \left(-t/\tau_{corr}\right)$, with $\tau_{corr}=0.15$ orbits. This
  dotted curve is seen to match nicely the numerical result.}
\label{S_function}
\end{center}
\end{figure}

\begin{figure*}
\begin{center}
\includegraphics[scale=0.6]{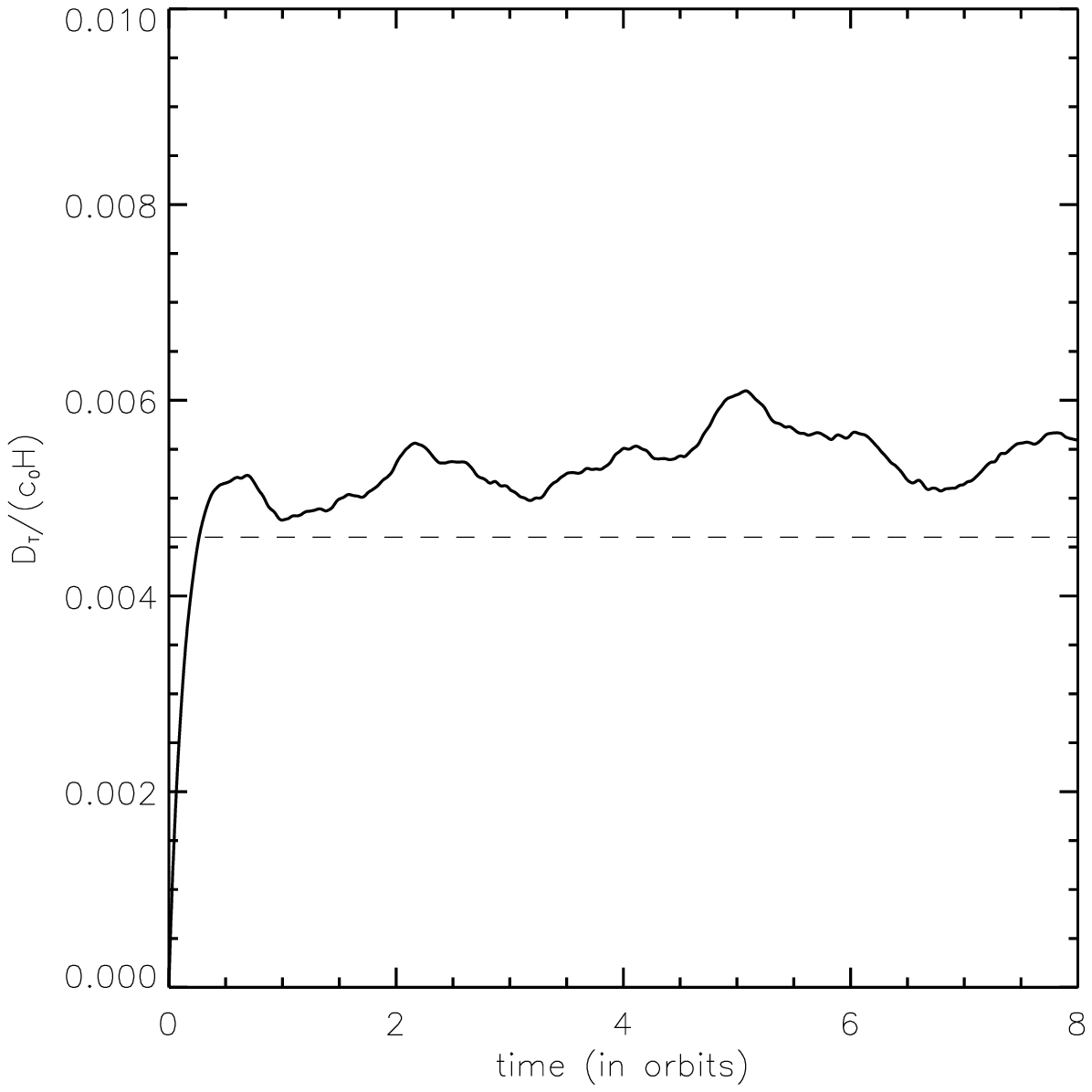}
\includegraphics[scale=0.6]{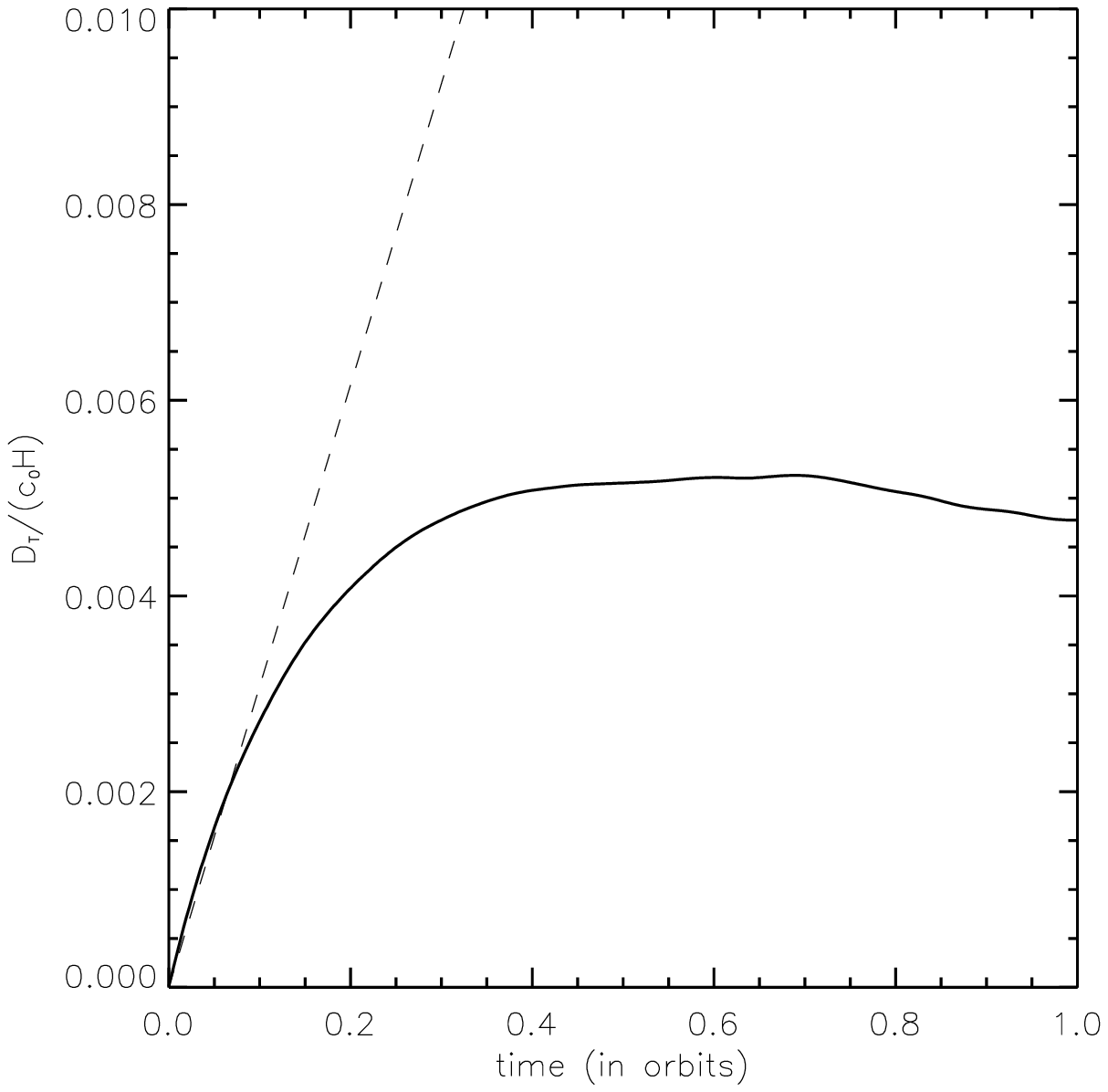}
\caption{Diffusion coefficient as a function of time, normalised by
  $c_sH$, obtained from the simulations done with ZEUS--3D. It is
  computed using equation~(\ref{diff_coeff_func}) and
  combining an average over a part of the computational box
  ($|z|<H$) and over a set of models (see the
  description in the text). As shown on the left panel, the diffusion
  coefficient is observed to saturate to a well defined value
  $D\sim5.5\times 10^{-3}c_sH$ after an initial rise. The dashed line
  plots the analytical estimate computed from
  equation~(\ref{large_tau}). The right panel
  shows an enlargement of the early evolution of $D_T$, compared with
  its expected early behaviour as given by equation~(\ref{small_tau}) for
  small time ({\it dashed line})}
\label{dust_coeff}
\end{center}
\end{figure*}

\begin{figure*}
\begin{center}
\includegraphics[scale=0.6]{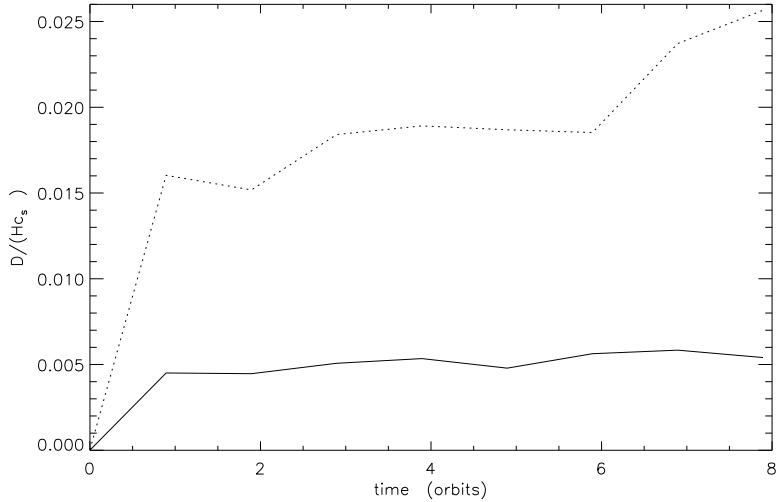}
\caption{As in Figure \ref{dust_coeff} but for runs performed
with NIRVANA. The lower curve applies for $|z| < H,$ and
the upper curve for $H < |z| < 2H .$ In the former case the diffusion
  coefficient is observed to attain 
  $D\sim 5\times 10^{-3}c_sH$ after an initial rise in good agreement
  with the ZEUS results. In the latter
  case the estimated diffusion coefficient is less stable but is about
  $4$ times larger.
  In that case the significance of the diffusion coefficient is less  clear
  for the reasons given in the text.}
\label{dust_coeffN}
\end{center}
\end{figure*}

In order to make a more quantitative comparison between the
results of these simulations and the simple theory presented in
section~\ref{simple_th}, we computed the functions $S_{zz}$ and $D_T$
according to equations~(\ref{Szz_func}) and (\ref{diff_coeff_func}). To
calculate the former, we first volume averaged the velocity product
in radius and azimuth. The vertical average is taken only within $H$
around the disk equatorial plane. Indeed, the behaviour  of the
turbulence away from the midplane is likely to depart very much from
being homogeneous and isotropic
 and may be influenced by the vertical boundary conditions.
 To reduce the statistical noise, the result
is then averaged over the nine different models we computed. The resulting
function $S_{zz}(\tau)$ is plotted in figure~\ref{S_function} using the
solid line. The dotted line represents the time variation of the function 
\begin{equation}
S_{zz}(\tau)=(0.07c_s)^2 e^{-t/\tau_{corr}} \, ,
\end{equation}
where $\tau_{corr}=0.15$ orbits. Given the good agreement between the
solid and dotted lines on figure~\ref{S_function}, we conclude that
$\tau_{corr}$ is a measure on the typical timescale over which the
fluid velocities become uncorrelated.

Integrating $S_{zz}$ over time gives $D_T(\tau)$. The left panel of
Figure~\ref{dust_coeff} shows the time evolution of $D_T/(c_s H)$.
Figure~\ref{dust_coeffN} gives results of similar calculations done
from runs with NIRVANA (see below).  $D_T$ is first observed to rise
at early times. This is in good agreement with
equation~(\ref{small_tau}). In fact, the prediction given by
equation~(\ref{small_tau}) is represented by the dashed line on the
right panel of figure~\ref{dust_coeff}, which is an enlargement of the
left panel at small times. It uses $(\delta v_z^2)^{1/2}=0.07 c_s$, as
obtained in section~\ref{model_prop}. This early linear rise of the
diffusion coefficient was also observed by
\citet{carballidoetal05}. Here, we see that it is naturally understood in
terms of the fluid velocity correlations. After this initial rise,
$D_T(\tau)$ is observed to reach a roughly constant value of $5.5 \times
10^{-3} c_sH$. This value nicely compares with the na\"ive estimate of
equation~(\ref{large_tau}). Indeed, taking the value of $\delta v_z^2$
and $\tau_{corr}$ derived above, one obtain $D_T \sim 4.6 \times
10^{-3} c_sH$, a value which is surprisingly close to the result of
the numerical simulations. This analytical estimate is represented by
the dashed line on figure~\ref{dust_coeff}.

\begin{figure*}
\begin{center}
\includegraphics[scale=0.85]{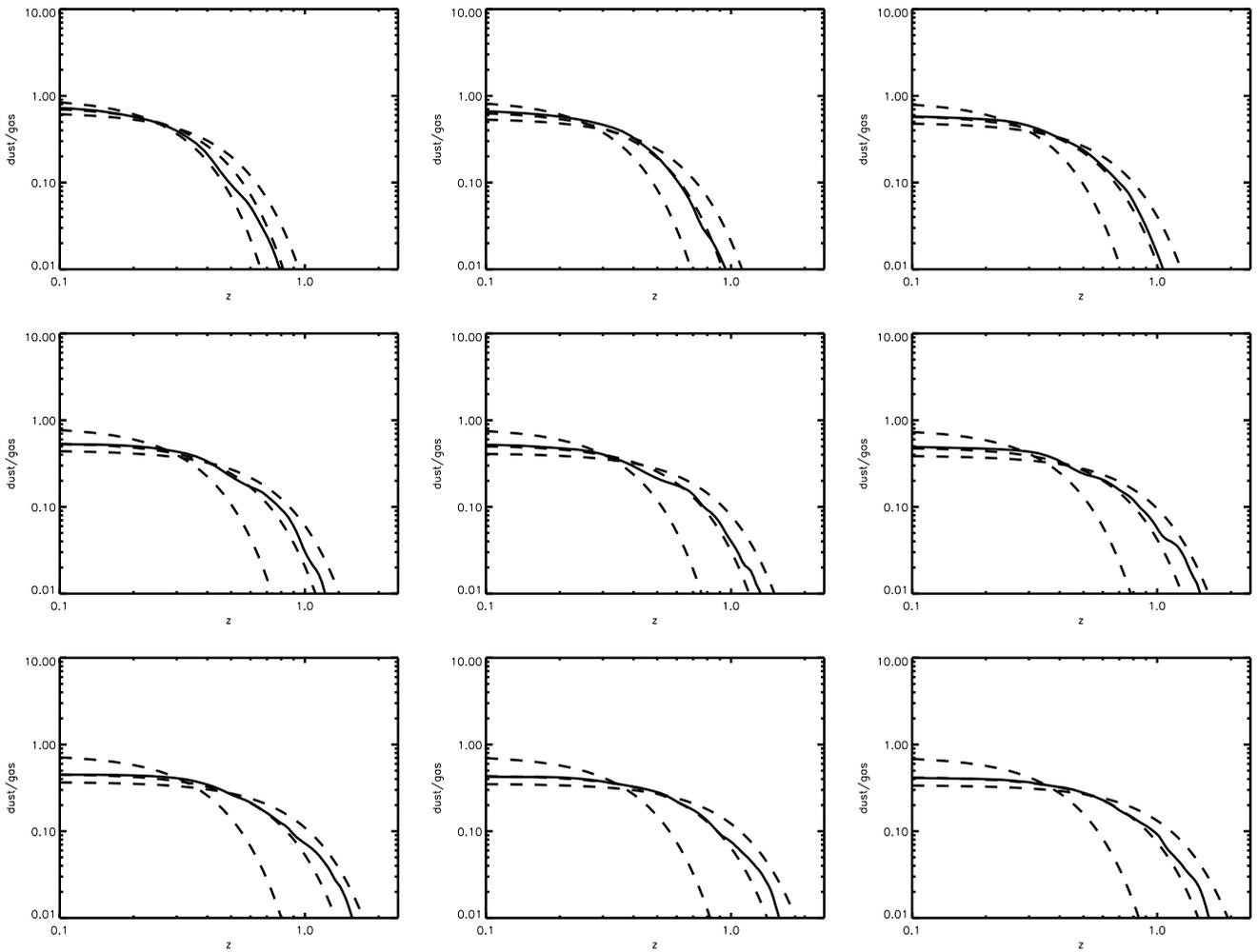}
\caption{Vertical profile of the dust to gas ratio as a function of
  time. The different panels correspond, from top to bottom and from
  left to right, to $t=0.48$, $0.80$, $1.12$, $1.44$, $1.76$, $2.08$,
  $2.40$, $2.72$ and $3.04$ orbits. On
  each panels, the solid line is the result of the simulations
  performed with ZEUS--3d,
  averaged over the nine models described in the text. Three dashed
  curves are plotted. They
  show the solution of equation~(\ref{diff_eq}), computed using
  $D/(c_s H)=10^{-3}$, $D/(c_sH)=5.5\times10^{-3}$ and $D/(c_sH)=10^{-2}$. The smaller the value
  of D, the smaller the diffusion.}
\label{diffusion_th}
\end{center}
\end{figure*}

If the dust distribution undergoes diffusive evolution as indicated
by the simple theory, $\rho_d$ should
satisfy  a diffusion equation of the form
\begin{equation}
\frac{\partial \rho_d}{\partial t}=D \frac{\partial}{\partial z}
  \left[ \rho \frac{\partial}{\partial z} \left( \frac{\rho_d}{\rho} \right) \right],
\label{diff_eq}
\end{equation}
where $D$ is a constant diffusion coefficient. We tested this
hypothesis by solving equation~(\ref{diff_eq}). To do so, we took an
initial dust distribution identical to that of the numerical
model and investigate three different values for the diffusion coefficient
$D$. First, we used $D/(c_sH)=D_{th}=5.5 \times 10^{-3}$, i.e. the value
given by our
simple theoretical model in terms of the velocity fluctuations of the
fluid. If this theory is correct, we expect this solution to be close
to the numerical solution. In order to check its accuracy, we also
computed the solution of equation~(\ref{diff_eq}) using
$D/(c_sH)=10^{-3}$ and $D/(c_sH)=10^{-2}$.

The comparison between the analytical solution of
equation~(\ref{diff_eq}) and the results of the numerical simulations
is shown in figure~\ref{diffusion_th}. It is illustrated by nine
panels. They correspond, from top left to bottom right, to times
$t=0.48$, $0.80$, $1.12$, $1.44$, $1.76$, $2.08$, $2.40$, $2.72$ and
$3.04$ orbits (measured after the dust was introduced into the
disk). On each panel, there are four curves. The solid line
shows the vertical profile of the dust--to--gas ratio, averaged
between the nine models we ran and normalised by the value in the
equatorial plane at $t=0$. The three dashed curves are
the solutions of equation~(\ref{diff_eq}) using the three diffusion
coefficients mentioned above. Of course, the smaller the value of $D$,
the less the dust is spread across the disk at a given time.

The agreement between the numerical results and the simple model is fairly
good. In all panels, the solid line is seen to have the best agreement
with the middle dashed curve, calculated using $D=D_T$. The
agreement is especially good at low altitude (typically $z<H$). This
was to be expected: first, this is the region in space where we
performed the volume average used to calculate 
the function $S_{zz}$. Second, the theory supposes that
the turbulence is homogeneous and isotropic. This results in only possibly reasonable
modelling  near the
midplane. The anisotropy of the turbulence is expected to increase
away from the midplane, as the density stratification becomes
stronger. From figure~\ref{diffusion_th}, there are some indications
that the estimated diffusion coefficient might increase with
height. Indeed, at late times, for
$z>H$, the solid line shows better agreement with the dashed curve
computed using $D/(c_sH)=10^{-2}$ than at lower altitudes, indicating that
dust particles are spread more efficiently than the simple theory
suggests. This is also supported by results of runs performed with NIRVANA
to test the simple diffusion theory. As with ZEUS--3D we considered
runs restarted from the basic disk model after $33, 40,$ and $60$
orbits. We evaluated the function
$D_T(\tau)$ by performing ensemble averages both over grid points
such that $|z| < H,$ and for grid points such that $H <|z| < 2H$
and then averaging over the models. Results are shown in Figure
\ref{dust_coeffN}. The indication is, in agreement with ZEUS, that for
$|z| < H,$ $D_T$ approaches $\sim 5\times 10^{-3}
c_s H,$ but for $H < |z| < 2H,$ $D_T$ is $\sim 4$ times larger.

\section{Vertical settling of larger particles}
\label{large_part}

\begin{figure}
\begin{center}
\includegraphics[scale=0.065]{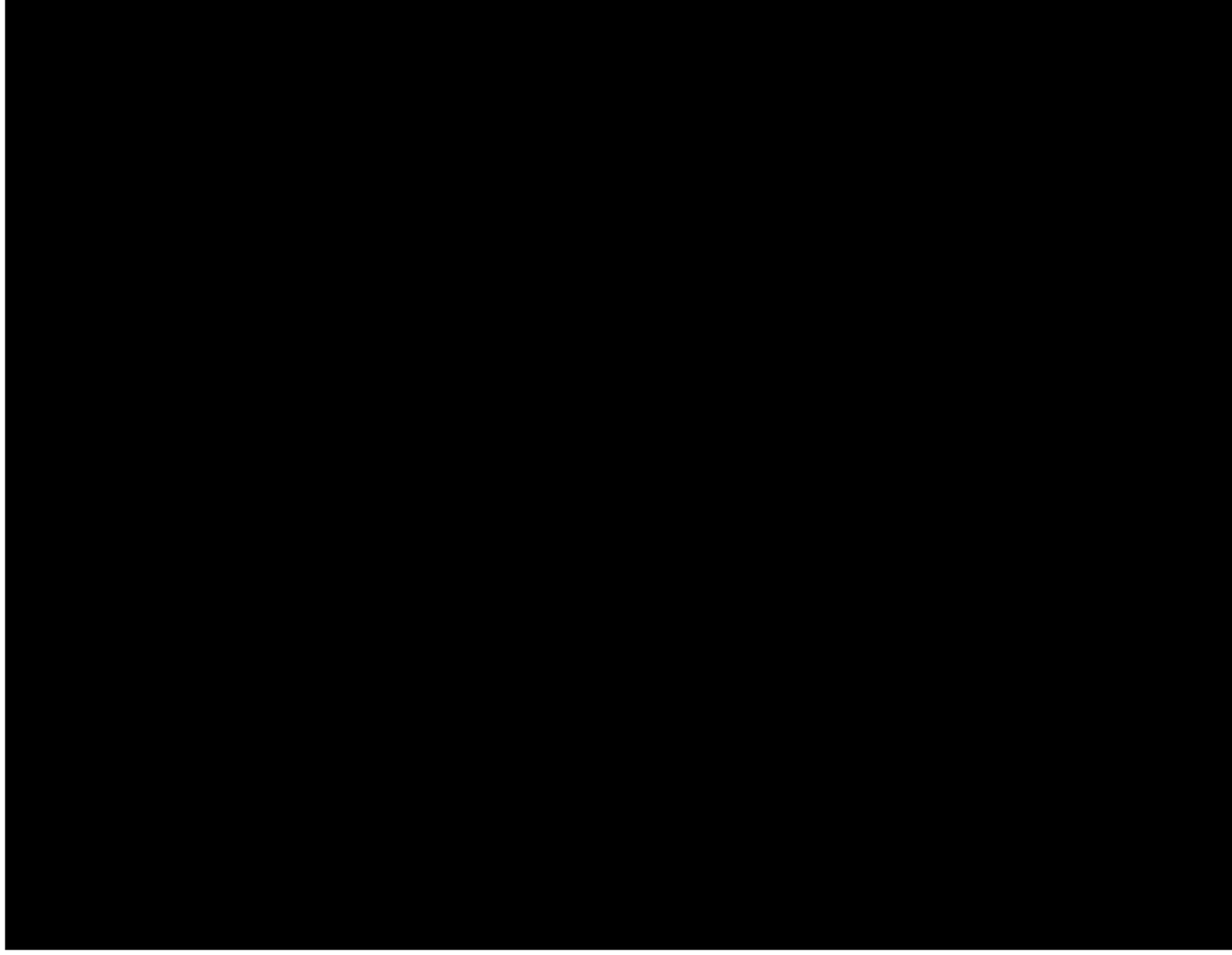}
\includegraphics[scale=0.065]{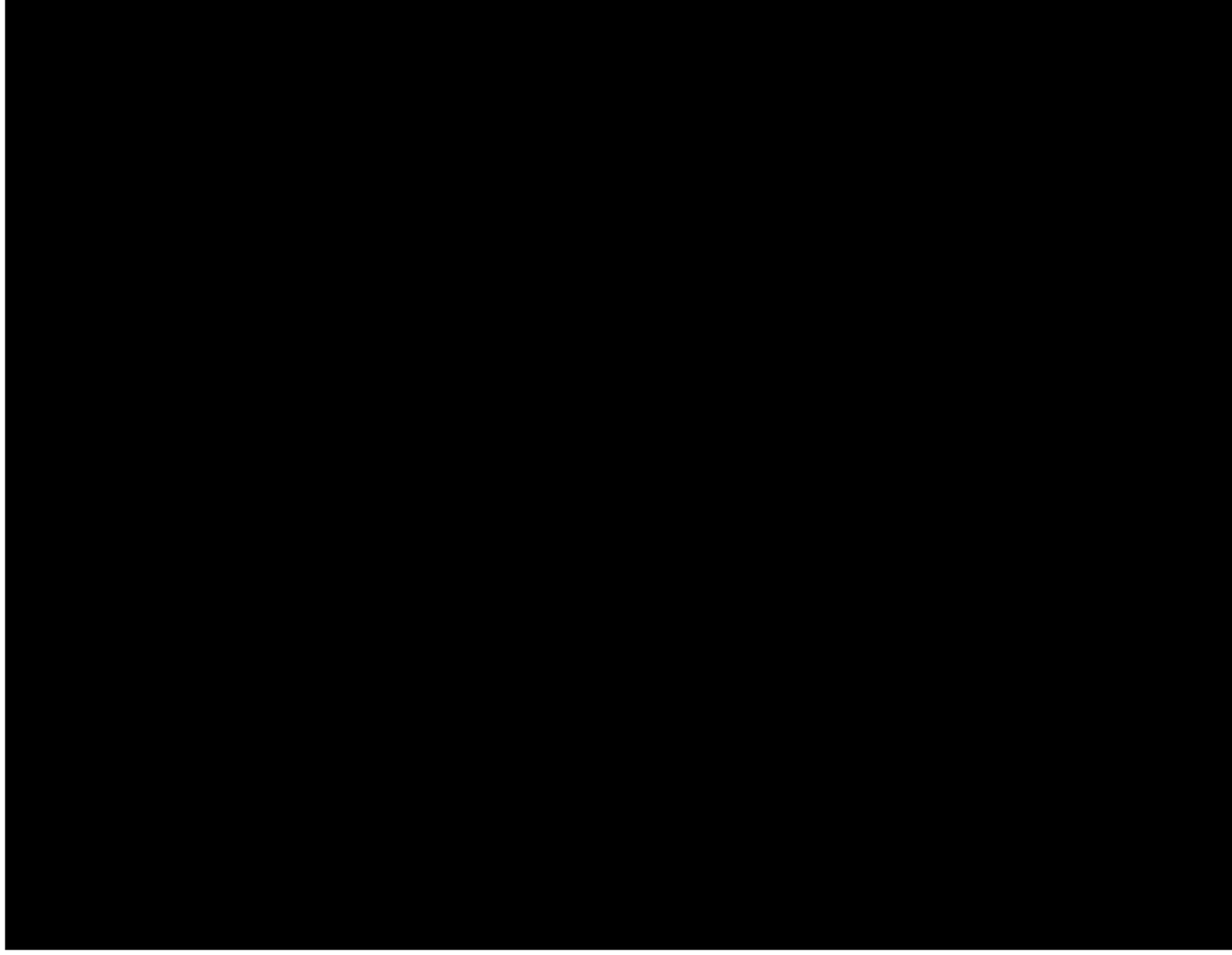}
\includegraphics[scale=0.065]{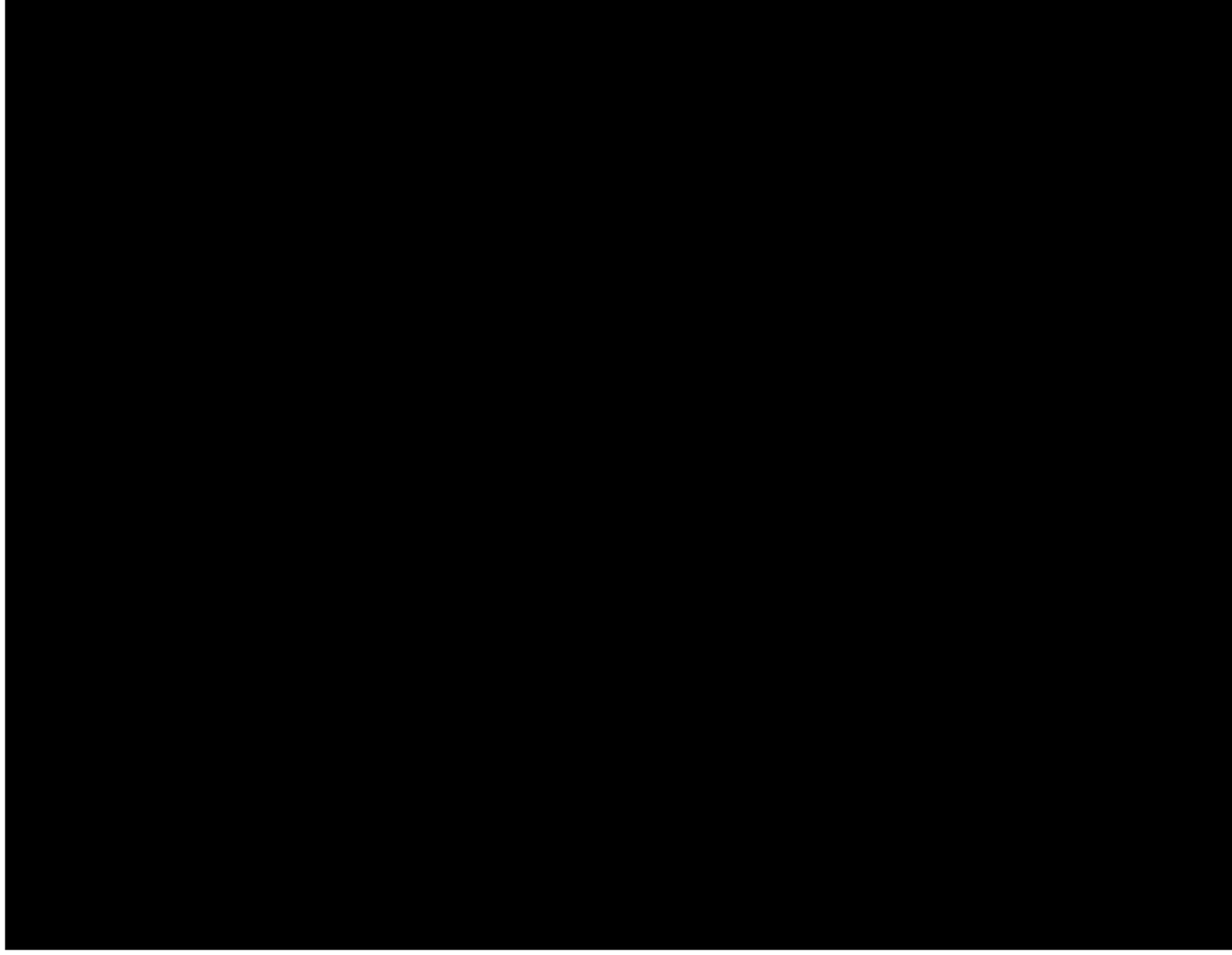}
\caption{Dust distribution in the (r,z) plane, obtained with ZEUS--3D,
  after a quasi--steady
  state has been reached. From left to right, the different panels
  correspond to $\Omega \tau_s=0.001$, $0.01$ and $0.1$.}
\label{larger_snapshot}
\end{center}
\end{figure}

\begin{figure}
\begin{center}
\includegraphics[scale=0.43]{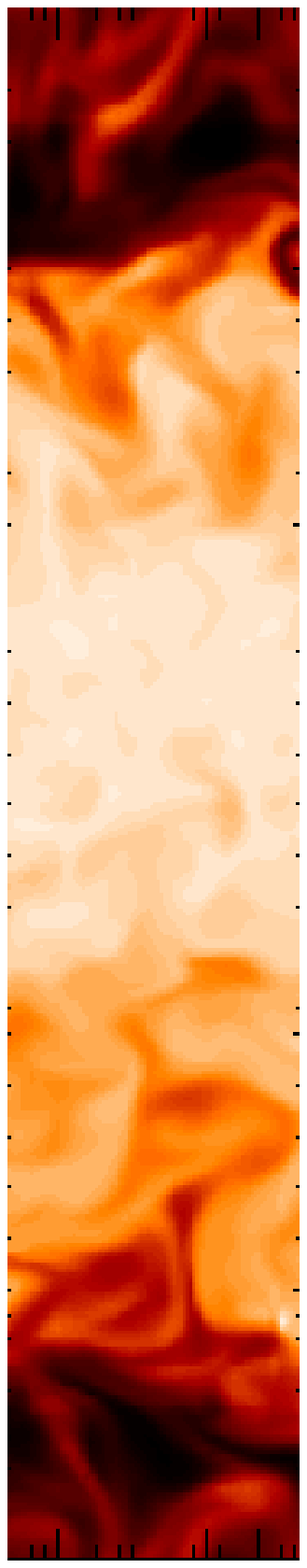}
\includegraphics[scale=0.43]{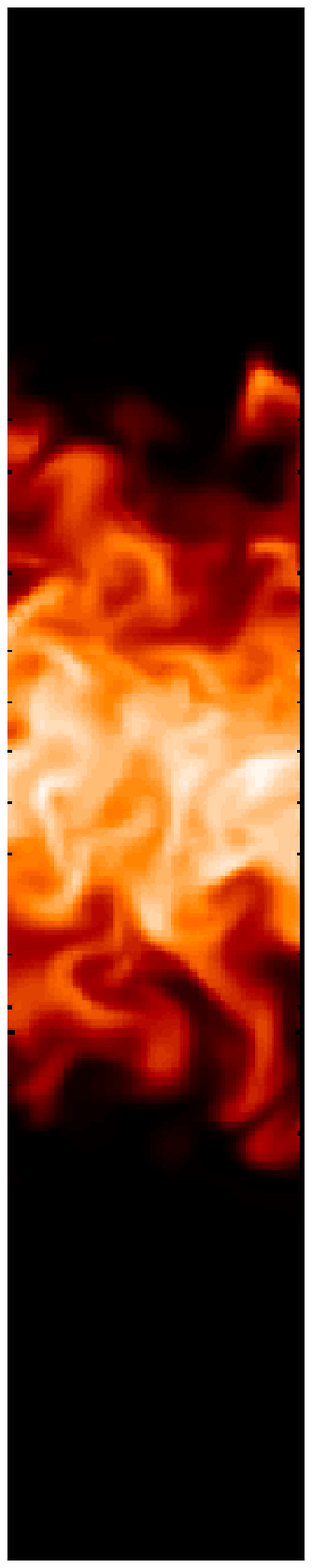}
\includegraphics[scale=0.43]{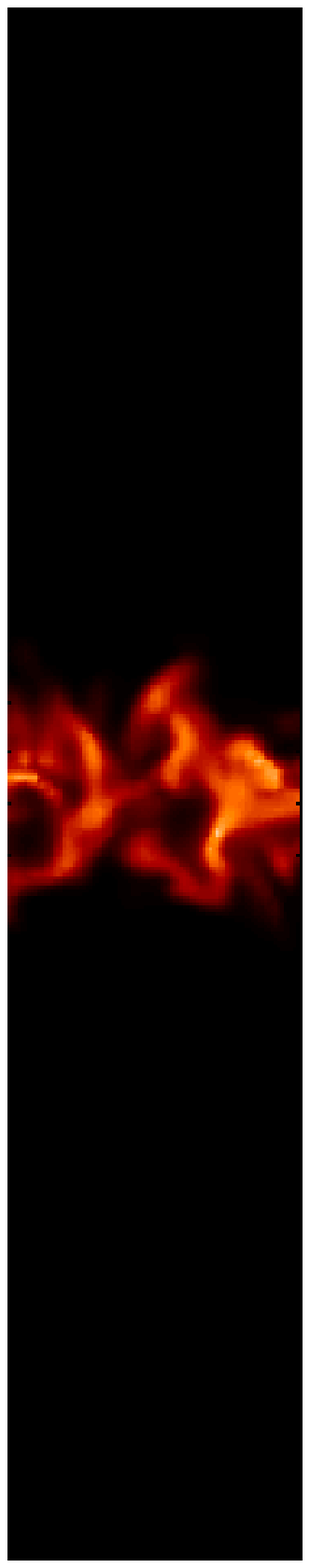}
\caption{As in Figure  \ref{larger_snapshot} but for results
obtained with NIRVANA}
\label{larger_snapshotN}
\end{center}
\end{figure}

\begin{figure}
\begin{center}
\includegraphics[scale=0.45]{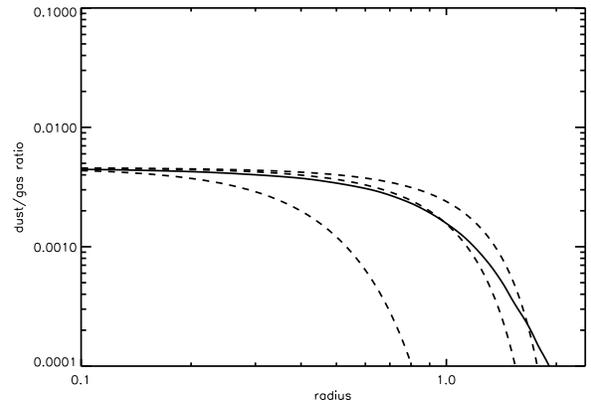}
\caption{Steady state vertical profile of the dust--to--gas ratio when
  $\Omega \tau_s=0.01$ ({\it solid line}). It has to be compared with
  the vertical profile calculated using equation~(\ref{diff_eq}),
  shown with the dotted line for three different values of the
  dimensionless diffusion coefficient $D/(c_sH)$: $10^{-3}$, $5.5\times10^{-3}$ and $10^{-2}$.}
\label{steady_1cm}
\end{center}
\end{figure}

\begin{figure}
\begin{center}
\includegraphics[scale=0.45]{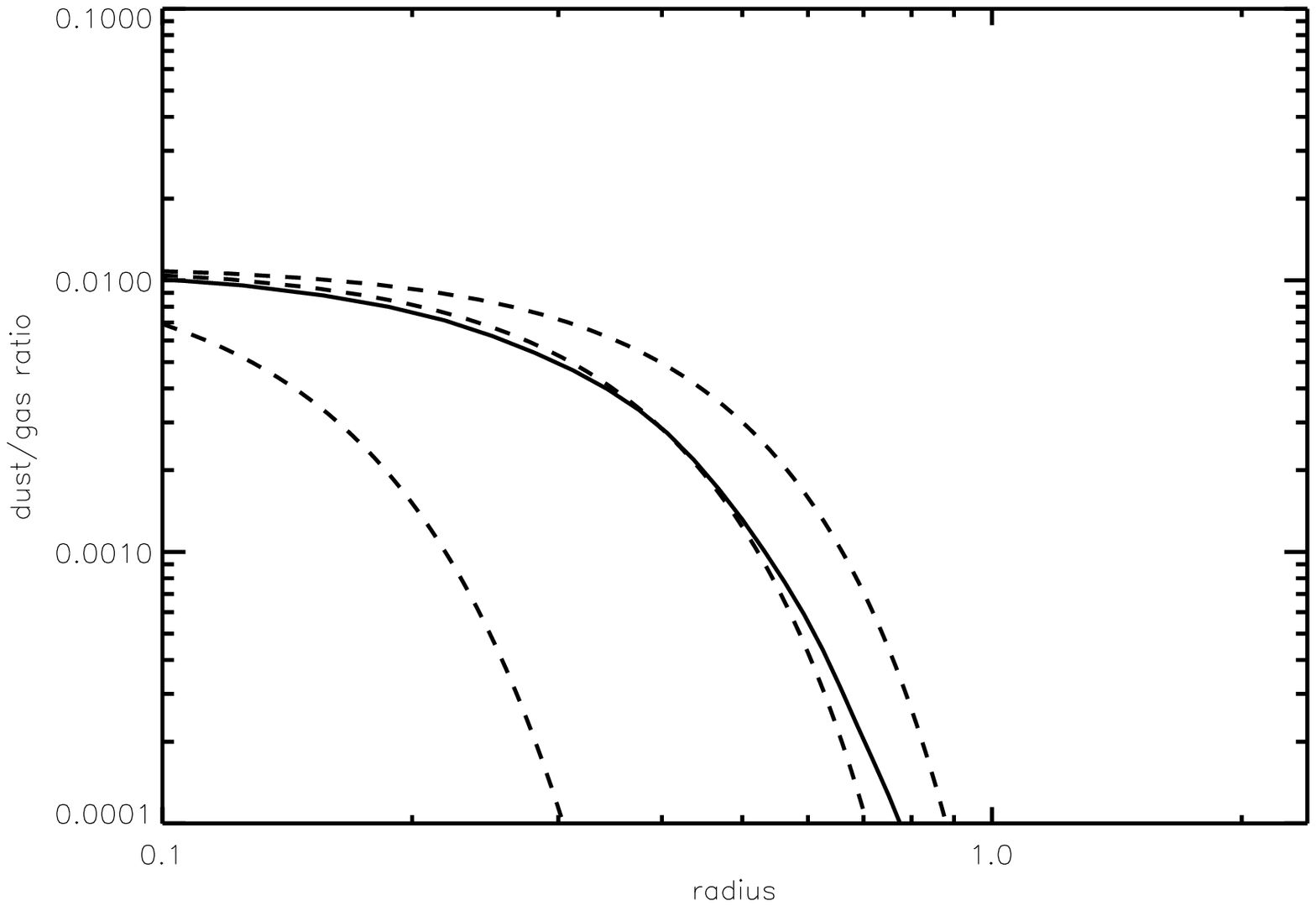}
\caption{Same as figure \ref{steady_1cm}, but for the case $\Omega
  \tau_s=0.1$.}
\label{steady_10cm}
\end{center}
\end{figure}

In the previous section, we have studied the effect of MHD turbulence
on very small dust particles, which behave essentially like  passive
scalars. In this section, we will focus on larger particles for which
the settling processes described in the introduction are fast in the
absence of turbulence.

We focus on three sizes, for which the parameter $\Omega \tau_s$ is
respectively $0.001$, $0.01$ and $0.1$ in the midplane of the initial
disk. Using equation~(\ref{size_part}), these values correspond respectively
to $1$ mm, $1$ cm and $10$ cm. In the absence of turbulence, the
typical settling timescale can be written \citep{dullemond&dominik04}
\begin{equation}
\frac{\tau_{sett}}{T_{orb}} \sim \frac{1}{2\pi} \frac{1}{\Omega
  \tau_s} \, ,
\end{equation}
where $T_{orb}$ is the orbital time. In our three cases, $\Omega
\tau_s=0.001$, $0.01$ and $0.1$  correspond to
$\tau_{sett}=160$, $16$ and $1.6$ orbits respectively. Thus
in the absence of turbulence, settling would
be important within the duration of the simulations in the last two cases.

At $t=40$ orbits, we introduced the particles into the turbulent disk
model with the
same vertical distribution as for the small particles discussed in
section~\ref{small_part}. Their evolution was then followed until the
vertical profile of the dust--to--gas ratio reaches a steady
state. Once again, by running models in which the initial
dust--to--gas ratio is uniform, we checked that this final
distribution does not depend on the initial conditions.

The results obtained with ZEUS--3D are illustrated in
figure~\ref{larger_snapshot}. This 
shows the typical distribution of the dust particles in the (r,z)
plane after they have reached a steady state. From left to right, the
snapshots corresponds to $\Omega \tau_s=0.001$, $0.01$ and
$0.1$. While the smallest dust particles are still efficiently spread
across the disk (compare with the fourth panel of
figure~\ref{micron_dust}), there are some signs of vertical settling
for the larger particles. This is obvious in particular for the
largest particles. But note however that turbulence is quite
efficient at preventing the complete settling of these
particles. As discussed above, a very thin dust subdisk would
form in just a few orbits in a quiescent disk. We have confirmed
that very similar results for the degree of settling
as a function of $\Omega \tau_s$ (evaluated in the midplane) are 
obtained with NIRVANA which employed
an advection diffusion treatment of the dust mass fraction.
Steady state distributions corresponding to those shown in
 Figure \ref{larger_snapshot} are plotted in  Figure \ref{larger_snapshotN}.
 Again these are found to be initial condition independent for distributions initiated
 in the midplane regions.

Once again, we can make use of the simple theory developed in
section~\ref{simple_th} to interpret these results. However, in this case the full
advection diffusion equation (\ref{advde}) incorporating  settling must be used
with an anomalous diffusion coefficient. Neglecting Lorentz forces, in the low $X$ 
or $\rho_d/ \rho$  limit
this gives \citep{dubrulleetal95,schapler&henning04}.  
\begin{equation}
\frac{\partial \rho_d}{\partial t}-\frac{\partial}{\partial z}(z
\Omega^2 \tau_s \rho_d)=D \frac{\partial}{\partial z} \left[ \rho 
\frac{\partial}{\partial z} \left( \frac{\rho_d}{\rho} \right)
 \right] \, .
\label{diff_advec_eq}
\end{equation}
Assuming a steady state, this equation gives a simple first order ordinary differential
equation for the vertical dust mass fraction profile. 
We calculate solutions for the same three  values of the diffusion coefficient 
that were used in
section~\ref{small_part} and compare them with the 
numerical simulations. The results are shown in figure~\ref{steady_1cm} for
the case  where $\Omega \tau_s=0.01$ and in figure~\ref{steady_10cm} for the
case $\Omega \tau_s=0.1$. We do not consider the case $\Omega \tau_s=0.001$
 because the dust--to--gas ratio is almost uniform at the end of
the simulation. In these figures the solid line shows the steady--state
dust--to--gas ratio obtained at the end of the simulations while the
three dashed curves are the corresponding  solutions of equation~(\ref{diff_advec_eq})
calculated with $D/(c_sH)=10^{-3}$, $5.5\times10^{-3}$ and
$10^{-2}$. As in section~\ref{small_part}, there is a good
agreement between the solid line and the middle dashed curve, for
which the magnitude of the diffusion coefficient 
corresponds to that  estimated using the velocity
fluctuations of the underlying  model. The two other values of the
diffusion coefficient can clearly be ruled out. There is also a marked 
difference between the solid and middle dashed curve in the upper
layers of the disk. This was also observed in
figure~\ref{diffusion_th}. As pointed out in section~\ref{small_part},
this is due to the increase of the diffusion coefficient at high disk
altitudes and also to the fact that our simple theory breaks
down because the turbulence ceases to be something that can
be profitably modelled as homogeneous and isotropic at
these locations.

\section{The effect of a dead zone}
\label{dead_zone}

All the results presented so far in this paper suppose that gas and
magnetic field are perfectly coupled throughout the entire vertical
extent of the disk. However, protoplanetary disks are probably 
cold and dense enough that this perfect coupling is unlikely, at least in some
regions of the disk. This situation has generated the ``layered accretion''
paradigm \citep{gammie96} in which the gas is turbulent only in the
upper layers of the disk, while a quiescent (or ``{\it dead}'') zone
exists around the midplane. The extent of this dead
zone is very uncertain, since it depends on the ionising source and on
the chemistry \citep{sanoetal00,fromang02,ilgner&nelson05}, but
its existence is likely. In this section, we investigate how
the presence of a dead zone would influence the results described in
the previous sections.

\begin{figure}
\begin{center}
\includegraphics[scale=0.5]{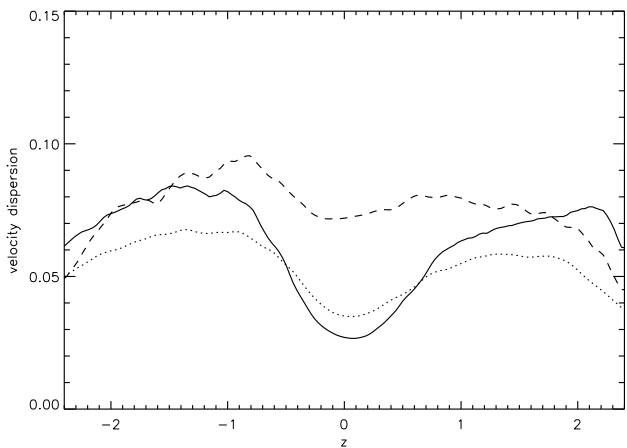}
\caption{Same as figure~\ref{vel_disp}, but for the ``larger dead
  zone'' model of \citet{fleming&stone03}.}
\label{vel_disp_dead}
\end{center}
\end{figure}

The problem of layered
turbulence  in local numerical simulations of stratified disks has
already been studied by \citet{fleming&stone03}. In this section, we
reproduce one of their  disk models by allowing the resistivity $\eta$ to be a
function of position. We choose the vertical profile of $\eta$
such that our model is identical to the ``larger dead
zone'' model of \citet{fleming&stone03}:
\begin{equation}
\eta=\eta_0 \exp \left(-\frac{z^2}{2}\right) \exp \left(
\frac{\Sigma_0}{\Sigma_{CR}} \frac{1}{2\sqrt{\pi}} \int_z^{\infty}
e^{-z'^2} dz'\right) \, ,
\end{equation}
where $\Sigma_0/\Sigma_{CR}=30$ and $\eta_0$ is chosen such that the
  Reynolds magnetic number $Re_m$, defined by
\begin{equation}
Re_m=\frac{c_sH}{\eta} \, ,
\end{equation}
equals $100$ in the midplane of the disk. This disk model was run 
for $100$ orbits. The evolution is the same as that  found by 
\citet{fleming&stone03}. In particular, we found that density waves are
excited in the dead zone by the turbulent motions in the active
layers. The velocity dispersions of the three velocity  components are
shown in figure~\ref{vel_disp_dead} as a function of $z$. This figure
should be compared with figure~\ref{vel_disp}.  Taken as a whole,
there is a decrease of the velocity
fluctuations compared to the fully turbulent model. The presence of
the dead zone is
 apparent through a decrease in the azimuthal and vertical velocity
fluctuations. The latter have a root mean square dispersion of roughly
$0.03 c_s$, thereby
showing a decrease by about a factor of two compared to the fully turbulent
model. However, since it does not vanish, a nonzero
diffusion coefficient can be expected.

\begin{figure}
\begin{center}
\includegraphics[scale=0.5]{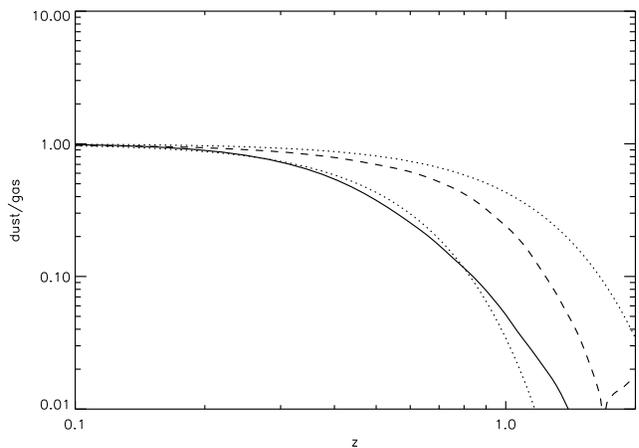}
\caption{Steady state vertical profile of the dust to gas ratio in
  the ideal MHD case ({\it dashed line}) and when a dead zone is
  present around the equatorial plane of the disk ({\it solid
  line}). The parameter $\Omega \tau_s$ equals $0.01$ in that case.}
\label{dead_zone_1cm}
\end{center}
\end{figure}

\begin{figure}
\begin{center}
\includegraphics[scale=0.5]{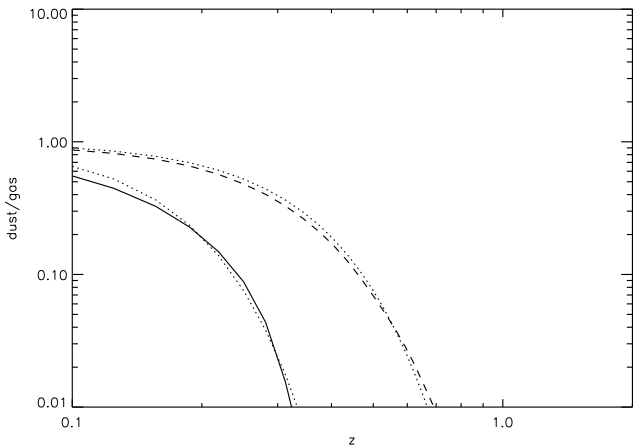}
\caption{Same as figure~\ref{dead_zone_1cm}, but for the case $\Omega
  \tau_s=0.1$.}
\label{dead_zone_10cm}
\end{center}
\end{figure}

Using this new underlying disk model and introducing dust particles, we used
ZEUS--3D to recalculate the models described  in
section~\ref{large_part} for which $\Omega \tau_s=0.01$ and
$\Omega \tau_s=0.1$. After a few orbits, the dust distribution reaches
a new equilibrium state. It is represented in
figure~\ref{dead_zone_1cm} (for $\Omega \tau_s=0.01$) and
figure~\ref{dead_zone_10cm} (for  $\Omega \tau_s=0.1$). In both cases,
the solid line shows the vertical profile of the dust--to--gas ratio in the
nonideal case with a dead zone, while the dashed line corresponds to the fully
turbulent case. As expected, the thickness of the dust
sub--disk is smaller in the former case.

It is possible to compare these results to  those obtained using the simple
model presented in section~\ref{simple_th}. To do so, we first note
that close to the midplane, the stopping time $\tau_s$ is nearly constant. A
steady--state solution to equation~(\ref{diff_advec_eq}) can be
written in that case as
\begin{equation}
\frac{\rho_d}{\rho}=\left( \frac{\rho_d}{\rho} \right)_0
e^{-z^2/2\tilde{H_d}^2} \, ,
\label{dust_profile}
\end{equation}
where the dust scale height $\tilde{H_d}$ is given by
\begin{equation}
\tilde{H_d}=\sqrt{\frac{D}{\Omega^2 \tau_s}} \, .
\label{dust_thickness}
\end{equation}
We first focus on the case $\Omega \tau_s=0.1$. When the disk is
completely turbulent (or {\it ``active''}), we found in
section~\ref{small_part} that the value of the diffusion coefficient
was $D/(c_sH)=5.5\times10^{-3}$. Using equation~(\ref{dust_thickness}),
this would give a scale height 
\begin{equation}
\tilde{H_d}^{active}=0.23
\end{equation}
for the dust sub--disk. Next, we seek an estimate of the dust
sub--disk scale height in the presence of a dead zone. By combining 
equation~(\ref{large_tau}) and equation~(\ref{dust_thickness}),
$\tilde{H_d}$ can be related to the velocity fluctuations by
\begin{equation}
\tilde{H_d} \propto (\delta v_z^2)^{1/2} \, .
\end{equation}
Given the smaller  value we obtained for $(\delta v_z^2)^{1/2}$ in the ``larger
dead zone'' model (see figure~\ref{vel_disp_dead}), we therefore expect
\begin{equation}
\tilde{H_d}^{dead} \sim \tilde{H_d}^{active}/2 \sim 0.11
\end{equation}
The two dust distributions computed with equation~(\ref{dust_profile})
using $\tilde{H_d}^{active}$ and
$\tilde{H_d}^{dead}$ (corresponding respectively to the fully turbulent
case and to the ``larger dead zone'' model) are plotted in
figure~\ref{dead_zone_10cm} using dotted lines. Both are seen to match
very accurately the solid and dashed curves that are their numerical
analogues.

The same procedure was followed in the model for which $\Omega
\tau_s=0.01$. In that case, we found $\tilde{H_d}^{active}=0.77$ and
$\tilde{H_d}^{dead}=0.37$. The dust--to--gas ratio vertical profiles
derived using these two values are plotted in
figure~\ref{dead_zone_1cm} with  dotted lines. The agreement with
the solid line is quite good. However, there is a poor agreement with
the dashed line. This is because the dust is spread to higher altitudes
in that case. The hypothesis that $\tau_s$ is a constant which we used
to derive equation~(\ref{dust_profile}) and (\ref{dust_thickness})
starts to  break down, which explain the discrepancy with the
numerical result.

\section{Discussion}
\label{discussion}

In this paper, we studied the effects of MHD turbulence on dust
settling by means on local numerical simulations
performed using ZEUS-3D and NIRVANA,  being Eulerian MHD codes using
finite differences.

We first investigated the case of very small particles which are strongly
coupled to the gas. Turbulent velocity fluctuations were found to
cause  an initially thin dust sub--disk to spread. The time evolution of the
vertical profile for the dust--to--gas ratio can be well modelled
by a diffusion equation, with a  diffusion coefficient $D$
that can  be expressed in terms of  turbulent  velocity correlations. We found that
a simple analytical estimate of $D$ can be obtained in terms of
 the  mean square amplitude of the
velocity fluctuations $\delta v_z^2$ and their correlation time
$\tau_{corr},$ both of which are properties of the turbulence alone:
\begin{equation}
D=(\delta v_z^2)^{1/2}\tau_{corr} \, .
\end{equation}
Similarly to \citet{turneretal06}, we found an increase of the
diffusion coefficient with disk height. We also noticed
that the ensemble averages used to calculate it show weaker
convergence when the upper layers of the disk are included. While the
diffusive description of dust spreading seems to work well in the neighbourhood of the
midplane, it is less accurate at disk heights exceeding  a few scale heights.

A standard procedure in this type of analysis is to determine the value of the
Schmidt number $S_c$, defined as the ratio between the anomalous viscosity
and the diffusion coefficient. The standard approach in dust diffusion
modelling is to take $S_c=1$
\citep{schapler&henning04,ilgneretal04,dullemond&dominik04}. In
non  zero net flux local simulations of radial dust diffusion,
\citet{carballidoetal05} found $S_c=11$, while \citet{johansen&klahr05}
reported $S_c=1.5$ for vertical diffusion in zero net flux
simulations. \citet{turneretal06} also reported a near unity Schmidt
number in their calculations. It is worth comparing these values to the
Schmidt number we can derive from our simulations. By averaging
the total stress represented in figure \ref{stresses_history} between
$20$ and $100$ orbits, one obtains $\alpha=1.54 \times
10^{-2}$. This, together with the value of the diffusion coefficient
obtained from the velocity fluctuations gives
\begin{equation}
S_c=\frac{\alpha c_s H}{D}=2.8 \, ,
\end{equation}
an intermediate value between the measures of \citet{johansenetal05}
and \citet{carballidoetal05}. However, it is important to stress here
that the
origin  of a non zero  diffusion coefficient is  on account of  the velocity fluctuations
and not in the transport properties of  angular momentum.

When dust particles grow to centimeter sizes, we found that they
start to decouple from the turbulence and settle towards the
midplane. The steady state profile of the
dust--to--gas ratio is well approximated by the solution of an
advection--diffusion  equation. Even for particles as
large as $10$ cm, we found that the dust sub--disk is significantly  spread
since its semi--thickness $H_d$ equals $0.23H$, while the settling
timescale in a quiescent disk is very short in that case ($1.6$
orbits). We note however that radial migration is important for
particles of this size and the interplay between that migration and
MHD turbulence in a stratified disk could lead to complex  phenomena, such as local
enhancement of the dust density
\citep{fromang&nelson05} that might affect this picture.

Because they are cold and dense,
 protoplanetary disks are unlikely to have adequate ionisation
 to be turbulent everywhere.
Therefore  we also investigated the effect of
the presence of a dead zone around the midplane. As expected, we found
thinner dust sub--disks in that case. Similarly to previous studies
\citep{fleming&stone03}, we found the dead zone is able to maintain 
significant activity (excited by the turbulent velocity
fluctuations of the active zone). This activity is able to prevent the
complete
settling of $10$ cm size particles. However, we want to emphasise 
that for computational reasons, our analysis was limited to a
case in which the mass of the dead zone roughly equals the mass of the
active zone. We expect our result to be modified in cases
where the mass of the dead zone is much larger than that of the active
zone and therefore only apply them to  dead zones
that involve a modest fraction of the local surface density.

Nonetheless for conditions appropriate to a minimum mass solar nebula,
the work presented here that considered MHD turbulence, taken together
with that of eg. \citet{gomez&ostriker05} indicates that gravitational 
instability of the dust layer is unlikely and that the formation of objects
of planetesimal size may depend on phenomena such as densification
in vortices operating together with vertical settling.

For practical reasons, we neglected grain growth in this work. This is
an important simplification, as dust particles are likely to grow at
the same time as they settle toward the equatorial plane of the disk
\citep{cuzzietal96,dullemond&dominik05}. Simulations of the evolution
of an entire
dust population through grain growth and turbulent stirring, that
take account of both radial and vertical disk structure,  are very
challenging with present day computational capabilities, but will have
to be performed in the future.

\section*{ACKNOWLEDGMENTS}
Some of the simulations presented in this paper were performed on the
QMUL High Performance Computing Facility purchased under the SRIF initiative.

\bibliographystyle{aa}
\bibliography{author}

\end{document}